# Magnetic superspace groups and symmetry constraints in incommensurate magnetic phases

**J. M. Perez-Mato[1], J. L. Ribeiro[2], V. Petricek[3] and M.I. Aroyo[1]**

[1] Dpto. de Física de la Materia Condensada, Facultad de Ciencia y Tecnología, Universidad del País Vasco, UPV/EHU, Apartado 644, 48080 Bilbao, Spain
[2] Centro de Física da Universidade do Minho, 4710-057 Braga, Portugal
[3] Institute of Physics, Academy of Sciences of the Czech Republic v.v.i., Na Slovance 2, 18221 Praha 8, Czech Republic

E-mail: *jm.perez-mato@ehu.es*

**Abstract**

Superspace symmetry is since many years the standard approach for the analysis of non-magnetic modulated crystals because of its robust and efficient treatment of the structural constraints present in incommensurate phases. For incommensurate magnetic phases, this generalized symmetry formalism can play a similar role. In this context we review from a practical viewpoint the superspace formalism particularized to magnetic incommensurate phases. We analyze in detail the relation between the description using superspace symmetry and the representation method. Important general rules on the symmetry of magnetic incommensurate modulations with a single propagation vector are derived. The power and efficiency of the method is illustrated with various examples, including some multiferroic materials. We show that the concept of superspace symmetry provides a simple, efficient and systematic way to characterize the symmetry and rationalize the structural and physical properties of incommensurate magnetic materials. This is especially relevant when the properties of incommensurate multiferroics are investigated.

## 1. Introduction

The use of the superspace formalism for the description of incommensurate modulated magnetic structures has already been proposed at the early stages of its development, more than thirty years ago [1]. However, although this theory has become the standard approach for the analysis of incommensurate and commensurate non-magnetic modulated crystals and quasicrystals [2,3,4], it has remained essentially unexplored as a practical approach to deal with magnetic incommensurate structures, except for some testimonial works [5]. This contrasts with the fact that incommensurate magnetic orderings are frequently found in magnetic phases, where the lattice geometry and the competitions between different types of interactions often leads to complex phase diagrams that include periodic and aperiodic (incommensurate) modulated phases [6,7]. But recently, the refinement program JANA2006 has been extended to deal with magnetic structures [8,9], and this code can now determine incommensurate magnetic structures using refinement parameters and symmetry constraints consistent with any magnetic superspace group. As a result, incommensurate magnetic phases have started to be investigated with the help of the superspace formalism as an alternative to the usual representation method [10-12].

The slower adoption of the superspace formalism in the case of magnetic incommensurate phases is related with the widespread use of the representation analysis developed by Bertaut [13,14]. This method is based on the decomposition of the magnetic configuration space into basis modes transforming according to different physically irreducible representations (irreps) of the space group of the paramagnetic phase (henceforth, *paramagnetic space group*), and it can be used to describe the magnetic modulations independently of their propagation vector being commensurate or incommensurate. The codes commonly employed for the refinement of incommensurate magnetic structures, such as FullProf [15], use this approach. However, this versatility has a cost. The recent upsurge of research work on multiferroic materials, where the spin-lattice coupling plays an essential role, has clearly shown both the limits of the representation method and the need for a comprehensive knowledge of how symmetry constrains the different magnetic and structural degrees of freedom and influences the physical properties of an incommensurate magnetic phase. This information is provided by the magnetic superspace formalism in a very simple and efficient manner [16,17]. For instance, the tensor properties of a given incommensurate phase are constrained by the magnetic point group of the magnetic superspace group assigned to that phase. In contrast, in the case of the representation method, the magnetic point group of the system is generally neither known nor controlled, and may even be inadvertently changed during the refinement, depending on the restrictions imposed on the basis functions.

The assignment of a superspace group symmetry to an incommensurate magnetic phase is therefore a fundamental step to rationalize its physical properties. As it happens for displacive modulations in non-magnetic incommensurate structures, a combined use of representation analysis and superspace formalism is highly recommendable [9]. The description of an incommensurate magnetic structure in terms of irrep modes is somewhat incomplete if the magnetic superspace group associated with the corresponding spin configuration is not explicitly given.

While for non-magnetic incommensurate structures the relationship between irrep modes and superspace formalism has been studied in detail [18-23], for magnetic structures it has only been recently considered for some specific materials [16,17]. To our knowledge a general practical framework for the combined use of the representation method and the superspace formalism in magnetic incommensurate phases has never been presented. The present article aims to fill this gap and draw the attention to the latter formalism by giving a comprehensive view on the application of the superspace symmetry concepts to magnetic incommensurate structures. After a brief review of the basic concepts of the superspace formalism, we will discuss in some detail the relationship between superspace symmetry and representation analysis. The power and efficiency of adopting the superspace description will then be illustrated through the analysis of several examples. For the sake of simplicity, and also because it is the most common case in modulated magnetic structures, we will restrict the discussion to systems with one-dimensional modulations, i.e. with a single propagation vector, for which the superspace has (3+1) dimensions.

## 2. Superspace symmetry and magnetic modulations

*2.1 Review of the basic concepts*

A complete and detailed introduction to the concepts of the superspace formalism can be found in the general references [2-4]. Here, we summarize the main results taking care that the arguments and the expressions explicitly include the case of magnetic structures.

A modulated magnetic structure with a single incommensurate propagation vector $\boldsymbol{k}$ is described within the superspace formalism by a normal periodic structure (the so-called basic structure, which has a symmetry given by a conventional magnetic space group $\Omega_b$), plus a set of atomic modulation functions defining the deviations from this basic periodicity of each atom in each unit cell. The magnetic space group $\Omega_b$ will be in general a subgroup of a paramagnetic space group. The modulation functions may concern the atomic positions, the magnetic moments, the thermal displacement tensor, some occupation probability or any other relevant local physical magnitude. The value of a property $A_\mu$ of an atom $\mu$ in the unit cell of the basic structure varies from one cell to another according to a modulation function $A_\mu(x_4)$ of period one, such that its value $A_{l\mu}$ for the atom $\mu$ at the unit cell $l$, with basic position $\boldsymbol{r}_{l\mu} = \boldsymbol{l} + \boldsymbol{r}_\mu$ ($\boldsymbol{l}$ being a lattice translation of the basic structure) is given by the value of the function $A_\mu(x_4)$ at $x_4 = \boldsymbol{k} \cdot \boldsymbol{r}_{l\mu}$:

$$A_{l\mu} = A_\mu(x_4 = \boldsymbol{k} \cdot \boldsymbol{r}_{l\mu}) \tag{1}$$

These atomic modulation functions can be expressed by a Fourier series of the type:

$$A_\mu(x_4) = A_{\mu,0} + \sum_{n=1,...} [A_{\mu,ns} \sin(2\pi n x_4) + A_{\mu,nc} \cos(2\pi n x_4)] \qquad (2)$$

Thus, a basic conventional periodic structure, a modulation wave vector $\mathbf{k}$ and a set of periodic atomic modulations $A_\mu(x_4)$ for each atom in the basic in unit cell determine the aperiodic values of any local atomic quantity and completely describe the aperiodic crystal. Considering the definition of $x_4$, such a description does not apparently differ much from the usual approach of using basis functions (waves) transforming according to irreps of the paramagnetic space group [15]. However, fundamental differences appear when the symmetry properties are defined.

By definition, any operation $(\mathbf{R},\theta\,|\,\mathbf{t})$ of the magnetic space group $\Omega_b$ of the basic structure (with $\mathbf{R}$ being a point-group operation, $\theta$ being -1 or +1 depending if the operation includes time reversal or not, and $\mathbf{t}$ a translation in 3d real space) transforms the incommensurately modulated structure into a distinguishable incommensurate modulated structure, sharing the same basic structure and having all its modulation functions changed by a common translation of the internal coordinate $x_4$, such that the new modulation functions $A'_\mu(x_4)$ of the $(\mathbf{R},\theta\,|\,\mathbf{t})$-transformed structure satisfy

$$A'_\mu(x_4) = A_\mu(x_4 + \tau) \qquad (3).$$

The translation $\tau$ depends on each specific operation. This implies that the original modulated structure can be recuperated by performing an additional translation $\tau$ along the so-called internal coordinate, i.e. the phase of the modulation functions. In this sense, one can speak of $(\mathbf{R},\theta\,|\,\mathbf{t},\tau)$ as a symmetry operation of the system defined in a four-dimensional mathematical space, where the fourth dimension corresponds to the continuous argument of the periodic modulation functions.

The addition of the global phase translation of the modulation as a fourth dimension allowing an additional type of transformations of the structure is enabled by the fact that an *arbitrary* phase translation of the modulation in an incommensurate phase (corresponding to the well-known phason excitations characteristic of incommensurate structures) keeps the energy invariant, in the same way that *arbitrary* rotations, roto-inversions, translations, and time reversal do. A symmetry group of a system is in general a subgroup of the group of transformations that keep the energy of the system invariant, and it is constituted by the operations of this group that have the additional property of leaving the system undistinguishable. Thus, space groups of commensurate structures are subgroups of the whole group of rotations, roto-inversions and translations. Similarly, in the case of an incommensurate structure, the symmetry group (the so-called superspace group) is defined as a subgroup of the full group of all transformations that keep the energy of the system invariant, including global arbitrary phase shifts of the incommensurate

Magnetic superspace groups and incommensurate magnetic phases modulation. The superspace group symmetry is then formed by the subset of $(R,\theta\,|\,t,\tau)$ operations that in addition keep the system undistinguishable after the transformation. The energy invariance for global phase translations therefore ensures the robustness of this generalized symmetry concept for characterizing the symmetry restrictions associated with an incommensurate phase [24]. It implies that the generalized symmetry, so defined, is a property that can be assigned to a thermodynamic phase and the break of this symmetry can only happen through a phase transition.

If $(R,\theta\,|\,t,\tau)$ belongs to the (3+1)-dim superspace group of an incommensurate magnetic phase, the action of $R$ on its propagation vector $k$ necessarily transforms this vector into a vector equivalent to either $k$ or $-k$. This means:

$$k \cdot R = R_I k + H_R \,, \tag{4}$$

where $R_I$ is either +1 or -1 and $H_R$ is a reciprocal lattice vector of the basic structure that depends on the operation $R$. The vectors $H_R$ can only be different from zero if the propagation vector $k$ includes a commensurate component [2].

The restrictions on the form of the atomic modulation functions that result from a superspace group operation $(R,\theta\,|\,t,\tau)$ can be derived from the above definitions as follows. If in the basic structure an atom $\nu$ is related with an atom $\mu$ by the operation $(R,\theta\,|\,t)$ such that $(R\,|\,t)r_\nu = r_\mu + l$, then their atomic modulation functions are not independent and are related by:

$$A_\mu(R_I x_4 + \tau_o + H_R \cdot r_\nu) = Transf(R,\theta) A_\nu(x_4)\,, \tag{5}$$

where $\tau_o = \tau + k \cdot t$ and $Transf(R,\theta)$ is the operator associated with the transformation of the local quantity $A_\mu$ under the action of the point-group operation $(R,\theta)$. Thus, equation (5) introduces a relationship between the modulation functions of the magnetic moments of the two atoms:

$$M_\mu(R_I x_4 + \tau_o + H_R \cdot r_\nu) = \theta \det(R) R \cdot M_\nu(x_4)\,, \tag{6}$$

while the atomic modulation functions $u_\mu(x_4)$, $u_\nu(x_4)$ defining the atomic displacements in each basic cell with respect to the basic positions $r_{l\mu}$ and $r_{l\nu}$ relate as:

$$u_\mu(R_I x_4 + \tau_o + H_R \cdot r_\nu) = R \cdot u_\nu(x_4) \tag{7}$$

These relations imply that only the modulation functions of the set of atoms in the asymmetric unit of the basic structure are necessary in order to define the whole structure. Notice that equations (6) and (7) force specific restrictions on the possible forms of the modulation functions of atoms that occupy positions in the basic structure that are left invariant ($\mu = \nu$) by some symmetry operations of $\Omega_b$.

According to the above definitions, all translations of the basic lattice combined with conveniently chosen phase shifts, namely the operations $(I,+1|t,-k.t)$ (here, $I$ represents the identity matrix), belong to the superspace group of the structure and form its (3+1)-dim lattice. If $k = (k_x, k_y, k_z)$ is expressed in the basis reciprocal to the chosen direct basis of space group $\Omega_b$, then the four elementary translations $(I,+1|1\,0\,0,-k_x)$, $(I,+1|0\,1\,0,-k_y)$, $(I,+1|0\,0\,1,-k_z)$ and $(I,+1|0\,0\,0,1)$ generate that lattice and define a unit cell in the (3+1)-dim superspace. In the basis formed by these superspace unit cell translations, the symmetry operation $(R,\theta|t,\tau)$ can be expressed in the standard form of a space group operation in a 4-dim space, $(R_s,\theta|t_s)$, where $t_s$ is a four dimensional translation and $R_s$ a $4\times 4$ integer matrix defining the transformation of a generic point $(x_1, x_2, x_3, x_4)$:

$$R_S = \begin{pmatrix} R_{11} & R_{12} & R_{13} & 0 \\ R_{21} & R_{22} & R_{23} & 0 \\ R_{31} & R_{32} & R_{33} & 0 \\ H_{R1} & H_{R2} & H_{R3} & R_I \end{pmatrix} \tag{8}$$

Here, the $R_{ij}$ are the matrix coefficients of the rotational 3-dim operation $R$ of the space group operation $(R,\theta|t)$ belonging to $\Omega_b$ (expressed in the basis of the basic unit cell), $(H_{R1}, H_{R2}, H_{R3})$ the components (in the corresponding reciprocal basis) of the vector $H_R$ defined in (4), and $R_I$ is +1 or -1, according to equation (4). The superspace translation $t_s$ in the 4-dim basis is given by $(t_1, t_2, t_3, \tau_0)$, where $t_i$ are the three components of $t$ in the basis of the basic unit cell and $\tau_o = \tau + k \cdot t$, as in equation (5). The value of $\tau_0$ does not depend on the specific value of the irrational component(s) of the incommensurate wave vector $k$, and the group composition law is a trivial extension of the usual law for conventional (3dim) space groups:

$$(R_{s1},\theta_1|t_{s1})(R_{s2},\theta_2|t_{s2}) = (R_{s1}R_{s2}, \theta_1\theta_2|R_{s1}\cdot t_{s2} + t_{s1}) \tag{9}$$

Superspace groups can therefore be defined with symmetry cards entirely analogous to those of normal space groups (see example in 2.2 below).

A superspace group operation $(R_s,\theta\,|\,t_s)$ can be symbolically expressed in a generalized Seitz-type simpler form $\{R,\theta\,|\,t_s\}$, with $t_s = (t_1, t_2, t_3, \tau_0)$, only indicating explicitly the operation $R$ [since the $4\times 4$ matrix $R_s$ is fully determined by $R$ (see equation (8)) while keeping the translational part expressed in the superspace unit cell basis. We will use the keys { } to distinguish this form of expressing the superspace symmetry operations, which obviates the ever present $-k\cdot t$ internal translation along $x_4$. In the following, we will use when appropriate one or the other notation; their equivalence, $(R,\theta\,|\,t_1\,t_2\,t_3,\tau) = \{R,\theta\,|\,t_1\,t_2\,t_3\,\tau_o\}$ with $\tau_o = \tau + k\cdot t$, should be kept in mind. For instance $(R,\theta\,|\,0\,0\,\tfrac{1}{2},\tfrac{1}{2}-\tfrac{1}{2}\gamma)$ is the same as $\{R,\theta\,|\,0\,0\,\tfrac{1}{2}\,\tfrac{1}{2}\}$ (with $k=\gamma c^*$). In one case we are using the 3D translational lattice vectors of the basic structure, while in the other case we use the usual oblique lattice basis vectors of the superspace lattice.

Summarizing, an incommensurate magnetic structure can be fully described by specifying: *i*) its magnetic superspace group (as in normal crystallography, this symmetry group can be unambiguously given by listing its symmetry operations); *ii*) its periodic basic structure (usually non-magnetic), with its symmetry given by a conventional (magnetic) space group forced by the superspace group; and *iii*) a set of periodic atomic modulation functions (period 1) that define, according to equation (1), the magnetic modulations for the atoms of the asymmetric unit of the basic structure. If, besides the magnetic modulations, there exist additional structural modulations (as, for instance, lattice distortions induced by spin-lattice coupling), these will be described by their corresponding modulation functions defined for the atoms of the same asymmetric unit, and constrained by the same superspace group. The magnetic point group of the system is given by the set of all point-group operations present in the operations of this superpace group.

*2.2 The simplest example: a centrosymmetric incommensurate modulation*

Let us consider the simplest illustrative example: a paramagnetic phase with space group $P\overline{1}$ (magnetic group $P\overline{1}1'$) develops a magnetic modulation with an incommensurate propagation vector ($\alpha,\beta,\gamma$) directed along an arbitrary direction such that its superspace symmetry is given (besides the 4-dim lattice translations) by the representative operations: $\{1\,|\,0000\}$, $\{\overline{1}\,|\,0000\}$, $\{1'|000\tfrac{1}{2}\}$ and $\{\overline{1}'|000\tfrac{1}{2}\}$ [4]. This superspace group can be denoted as $P\overline{1}1'(\alpha\beta\gamma)0s$, using a natural extension of the well-established labelling rules for non-magnetic superspace groups [3,25], and as shown in section 3.3, it is the symmetry of any magnetic modulation originated by a single irreducible representation. Table 1 lists the symmetry operations of this group in the form of generalized symmetry cards, as used for instance in JANA2006 [8]; these cards use a self-explanatory notation, indicating unambiguously the linear transformations in the four-dimensional unit cell basis.

---

[4] Henceforth, when indicating concrete operations and not generic operations, we drop the index $\theta$ and indicate the inclusion of time reversal by adopting the usual convention of adding a prime to the point-group operation symbol (*1'*, $m_x'$, $2_y'$, ...).

**Table 1.** Representative operations of the centrosymmetric superspace group $P\bar{1}1'(\alpha\beta\gamma)0s$ described by using generalized Seitz type symbols (left column) and symmetry cards as used in the program JANA2006 [8].

| | | | | | |
|---|---|---|---|---|---|
| $\{1|0000\}$ | $x_1$ | $x_2$ | $x_3$ | $x_4$ | $+m$ |
| $\{\bar{1}|0000\}$ | $-x_1$ | $-x_2$ | $-x_3$ | $-x_4$ | $+m$ |
| $\{1'|000\frac{1}{2}\}$ | $x_1$ | $x_2$ | $x_3$ | $x_4 + \frac{1}{2}$ | $-m$ |
| $\{\bar{1}'|000\frac{1}{2}\}$ | $-x_1$ | $-x_2$ | $-x_3$ | $-x_4 + \frac{1}{2}$ | $-m$ |

Let us now see how these symmetry operations constrain the resulting magnetic and structural modulations. According to equations (6) and (7), the symmetry operation $\{1'|000\frac{1}{2}\}$ implies that the spin modulations $M_\mu(x_4)$ of all magnetic atoms must necessarily be odd functions for a translation ½. Therefore, their expansion is restricted to odd Fourier terms. Similarly, any induced structural modulations $u_\mu(x_4)$ that may occur as secondary effects are necessarily even for the same translation, and are therefore restricted to even Fourier terms. The inversion operation further restricts the modulations of atoms lying at special positions in the paramagnetic structure. According to (5) and (6), the Fourier series (see equation (2)) describing the magnetic modulations for atomic sites with inversion symmetry (Wyckoff positions 1a, 1b,…, 1h) can only have cosine terms, while the Fourier series of the induced structural modulations are restricted to sine terms. In addition, the modulation functions for an atom in a general position (*x,y,z*), say atom 1, determines the modulation of its symmetry related (*-x,-y,-z*) pair, say atom 2, according to the relations:

$$M_2(-x_4) = M_1(x_4) \qquad (10)$$

$$u_2(-x_4) = -u_1(x_4) \qquad (11)$$

These equations imply that the corresponding Fourier components must fulfil the conditions

Magnetic superspace groups and incommensurate magnetic phases

$M_{2,ns} = -M_{1,ns}$, $M_{2,nc} = M_{1,nc}$ (n-odd) and $u_{2,ns} = u_{1,ns}$, $u_{2,cs} = -u_{1,nc}$ (n-even), with the sub-indexes *s* and *c* indicating the sine and cosine Fourier amplitudes, respectively (see equation 2).

As the phase of the total modulation in an incommensurate phase is arbitrary, the above discussion restricting the modulations of atoms at centrosymmetric sites to cosine or sine terms can be misleading. In fact, the inversion operation for an arbitrary choice of this global phase of the modulation would be of the form $\{\bar{1}|000\tau\}$ with $\tau \neq 0$, but we have made a specific choice of this phase, equivalent to a choice of the origin in internal space, such that $\tau = 0$. Therefore, the important property, independent of the origin choice, is that the modulation functions for all atoms at special positions must necessarily be in phase (see figure 1), and that the possible induced structural modulations (which include only even Fourier terms) are necessarily shifted by $\frac{\pi}{2}$ or $-\frac{\pi}{2}$ with respect to the magnetic modulation.

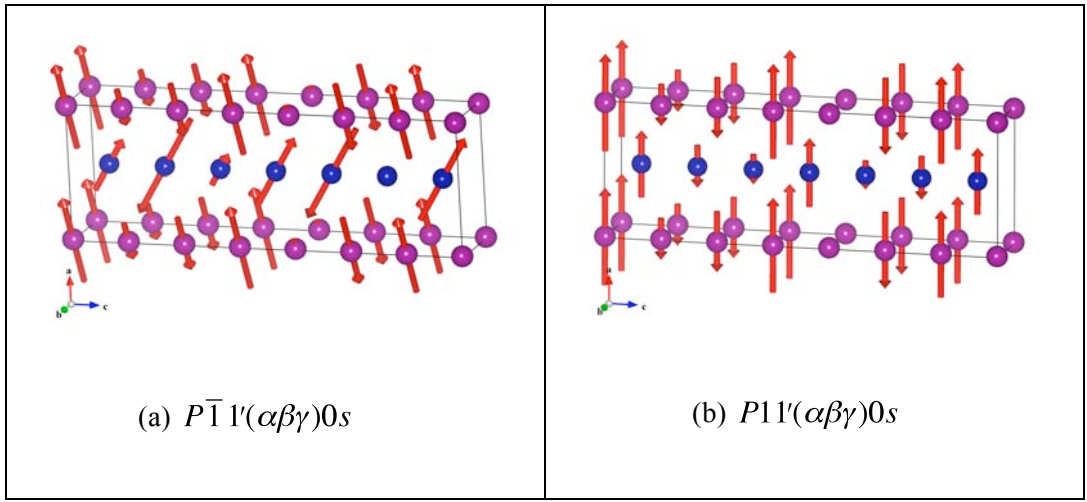

(a) $P\bar{1}1'(\alpha\beta\gamma)0s$  (b) $P11'(\alpha\beta\gamma)0s$

**Figure 1.** Examples of magnetic structures keeping (a) and breaking (b) centrosymmetry. In both cases, it is a triclinic $P\bar{1}$ structure with different magnetic atoms at sites (000) and (½ ½ ½) and propagation vector (0 0 ~0.32). They show that collinearity of all spin waves is not necessary for keeping the inversion centre. In case (a), the spin modulations in both independent atoms are in phase and the superspace group maintains the space inversion operation. In case (b), despite being collinear, the magnetic modulations of both atoms are phase shifted and the inversion symmetry is broken. The labels of the superspace symmetries corresponding to each case are indicated below.

The breaking of space inversion symmetry by an incommensurate modulation is sometimes difficult to visualize (see figure 1). If, for example, the system has several independent magnetic atomic sites in the paramagnetic phase, the restrictions that keep the inversion symmetry do not necessarily imply a collinear magnetic ordering. But the superspace formalism describes in a simple and general form both the structural and magnetic constraints associated with the presence of an inversion centre (see section 4 for more on this example).

## 3. Magnetic superspace groups and irreducible representations

In accordance with Landau theory, magnetic ordering is a symmetry-breaking process that can be described by an appropriate order parameter. In many cases, the transformation properties of this order parameter correspond to those of a single irreducible representation (irrep) of the *magnetic grey space group* associated with the paramagnetic phase. The frequent limitation of the magnetic modulation to a single propagation vector is often a consequence of this restriction to a unique irreducible order parameter, that is, an order parameter that is transformed according to a single irrep. In more general cases, magnetic configurations with a single propagation vector can be decomposed into several magnetic modes transforming according to different irreps sharing the same propagation vector. This is the basis for the representation analysis method developed by Bertaut [13,14], where the possible magnetic orderings are parameterized by complete sets of basis modes transforming according to the irreps of the paramagnetic space group associated with the observed propagation vector. The magnetic configuration is described with the help of basis modes corresponding to a single irrep or, if necessary, to a set of irreps as small as possible. It should be stressed that the irreps in this method are *ordinary* representations (but odd for time reversal) and these irreps define not only the transformation properties of the magnetic configuration for the operations keeping invariant the propagation vector $k$, but also for those transforming $k$ into $-k$. The introduction of corepresentations is therefore not necessary for dealing with these latter transformations (see also subsection 4.1.2).

It is important to establish in detail the relationship between the symmetry constraints imposed by the assignment of a certain irrep to the magnetic order parameter and those resulting from ascribing a magnetic superspace group to the magnetic phase. As we will see below, these two sets of constraints are closely related, but superspace symmetry is in general more restrictive and more comprehensive, as it affects all the degrees of freedom of the system.

*3.1. The parent symmetry for a magnetic modulated phase*

The parent symmetry to be considered for a magnetic phase is the magnetic grey space group of the paramagnetic phase. For each space group operation $(R\,|\,t)$, we have to consider two distinct operations $(R,-1\,|\,t)$ and $(R,+1\,|\,t)$, distinguishing if the operation $(R\,|\,t)$ is complemented by a time reversal transformation or not. The symmetry operations of the paramagnetic crystal are thus

Magnetic superspace groups and incommensurate magnetic phases

trivially doubled, implying that the magnetic space group $\Omega_p$ can be expressed as the coset expansion:

$$\Omega_p = G_p + (1'|0\,0\,0)G_p \qquad (12),$$

where $G_p$ is the ordinary space group formed by operations of type $(R,+1|t)$. The coset $(1'|0\,0\,0)G_p$ includes an equal number of operations $(R,-1|t)$. Notice that the antiunitary properties of the operations that integrate this second coset [26] are irrelevant when working with real quantities. Therefore, we do not need to use the co-representations of the grey group $\Omega_p$ in order to describe the transformation properties of a given arrangement of magnetic moments or atomic displacements. The irreps of $\Omega_p$ are trivially related with those of $G_p$: for each irrep of $G_p$ two irreps of $\Omega_p$ exist, one associating the identity matrix $\mathbf{1}$ to time reversal, and the other the matrix $-\mathbf{1}$. In the following, we shall call them *non-magnetic* and *magnetic* irreps, with explicit generic labels *T* and *mT*, following the notation employed in ISODISTORT [27]. As time reversal changes the sign of all magnetic moments, magnetic modes obviously transform according to magnetic irreps, while, phonon modes, for instance, transform according to non-magnetic irreps. The odd character for time reversal of the irreps of the magnetic modes is usually not explicitly indicated in conventional representation analyses of magnetic structures, but it is very important to make clear this distinction in a general context where lattice degrees of freedom are also classified according to irreps.

*3.2. The order parameter and the general invariance equation*

The components of the irreducible order parameter can be considered as amplitudes of a set of static spin waves with propagation vectors $\{k_1,....,k_n\}$ (the so-called wave vector star of the irrep) that transform into each other under the action of the symmetry group of the paramagnetic phase. If $N$ is the number of independent spin waves for the propagation vector $k_1$ (i.e., the dimension of the so called small representation), then there exist an equal number for all other wave vectors in the irrep star, and the dimension of the irrep is $n \times N$. In general, an incommensurate magnetic ordering with a single propagation vector and transforming according to a single irrep *mT* can give rise to different superspace group symmetries depending on the *direction* taken by the irrep order parameter in this $n \times N$ space. Notice that the term *irreducible representation* (irrep) is used here in the sense of *physically irreducible representation*, because we are concerned with the transformation properties of real physical magnitudes, such as magnetic moments or lattice distortions. Therefore, in some cases, these irreps are actually the direct sum of two complex

Magnetic superspace groups and incommensurate magnetic phases conjugate irreducible representations. This implies that the irrep star is always formed by pairs of wave vectors $k_i$ and $-k_i$.

Independently of the number of arms of the irrep star, the possible directions for the order parameter that yield a magnetic ordering with a single propagation vector (and therefore a symmetry described by a (3+1)-dim superspace group) are necessarily limited to those where only a single wave vector $k$ (and its opposite, $-k$) of the irrep star is involved. We can then constrain the order parameter to a *2N*-dim subspace within the irrep space, and express the magnetic moment $M(\mu,l)$ of any atom $(\mu,l)$ in the structure as:

$$M(\mu,l) = \sum_{i=1,...,N} S_i(k)\mathbf{m}_i(\mu)e^{-i2\pi k.(l+r_\mu)} + S_i(-k)\mathbf{m}_i^*(\mu)e^{i2\pi k.(l+r_\mu)} \qquad (13)$$

Here, $S_i(k)$ and $S_i(-k)$ are global complex components of the order parameter, (with $S_i(-k) = S_i^*(k)$, $i = 1,...,N$), $\mu$ labels the magnetic atoms in the reference unit cell and $\mathbf{m}_i(\mu)$ denotes a normalized *polarization vector* that defines the internal structure (i.e. the correlation between the atomic magnetic moments in a unit cell ) of each of the *N* spin waves. Notice that the choice for the sign of the exponents in equation (13) complies with the convention of a positive phase shift $e^{i2\pi k.t}$ for the action of a translation $(1|t)$ on the spin wave amplitudes $S_i(k)$ (see also equation (15) below). Notice also that we are defining a single global magnetic mode $\{\mathbf{m}_i(\mu)\}$ for each component of the order parameter. The magnetic moments $\mathbf{m}_i(\mu)$ of this mode will have correlations among symmetry related atoms according to the requirements of the transformation properties of the relevant irrep, along with specific physical correlations, as it is in general given by some system-dependent linear combination of basis modes with the same transformation properties.

By definition, an operation $(R,\theta|t)$ of the paramagnetic symmetry group such that $k \cdot R$ is equivalent either to $k$ or to $-k$, transforms any magnetic ordered configuration described by (13) with a set of amplitudes $\{S_i(k), S_i(-k)\}$ into a new one, described by the same equation and polarization vectors $\mathbf{m}_i(\mu)$ but with new transformed amplitudes $\{S'_i(k), S'_i(-k)\}$ given by:

$$\begin{pmatrix} S'_1(k) \\ ... \\ S'_N(k) \\ S'_1(-k) \\ ... \\ S'_N(-k) \end{pmatrix} = mT(R,\theta|t) \begin{pmatrix} S_1(k) \\ ... \\ S_N(k) \\ S_1(-k) \\ ... \\ S_N(-k) \end{pmatrix}. \qquad (14)$$

Magnetic superspace groups and incommensurate magnetic phases

Here, $mT(R,\theta|t)$ denotes a $2N \times 2N$ matrix that describes the operation $(R,\theta|t)$ within the $\{k,-k\}$ subspace of the irrep $mT$. For instance, in the simple case of a lattice translation $(1|t)$, the matrix $mT$ will be of the form:

$$mT(1|t) = \begin{pmatrix} \mathbf{1} \cdot e^{i2\pi k \cdot t} & \mathbf{0} \\ \mathbf{0} & \mathbf{1} \cdot e^{-i2\pi k \cdot t} \end{pmatrix} \quad (15)$$

with $\mathbf{1}$ and $\mathbf{0}$ representing the N-dimensional identity and null matrices, respectively. According to the definition of superspace symmetry introduced in section 2 (see equation 4), only the operations that keep the order parameter within this limited subspace of two opposite vectors ($k$ and $-k$) may be part of the superspace group. They form, in general, a subgroup of the paramagnetic grey space group that we shall call the *extended* little group of $k$ and denote by $\Omega_{p,k,-k}$. If the paramagnetic grey space group is non-polar, the extended little group $\Omega_{p,k,-k}$ can always be decomposed into two cosets:

$$\Omega_{p,k,-k} = \Omega_{p,k} + g_{-k}\Omega_{p,k}, \quad (16)$$

with $\Omega_{p,k}$ being the so-called little group that includes all operations keeping $k$ invariant (up to a reciprocal lattice translation), while the coset $g_{-k}\Omega_{p,k}$ includes an equal number of operations transforming $k$ into $-k$. If the grey group is a polar group, the second coset may not exist, in which case the extended little group coincides with the little group $\Omega_{p,k}$.

A phase shift $\alpha$ of the spin wave (see section 2) simply adds a phase factor to the amplitudes of the order parameter, transforming $\{S_i(k), S_i(-k)\}$ into $\{e^{i\alpha}S_i(k), e^{-i\alpha}S_i(-k)\}$. Therefore, a superspace operation $(R,\theta|t,\tau)$ exists for a spin configuration $\{S_i(k), S_i(-k)\}$ described by equation (13), if there is a real value $\tau$ such that:

$$\begin{pmatrix} \mathbf{S(k)} \\ \mathbf{S(-k)} \end{pmatrix} = \begin{pmatrix} \mathbf{1} \cdot e^{i2\pi\tau} & \mathbf{0} \\ \mathbf{0} & \mathbf{1} \cdot e^{-i2\pi\tau} \end{pmatrix} \cdot mT(R,\theta|t) \cdot \begin{pmatrix} \mathbf{S(k)} \\ \mathbf{S(-k)} \end{pmatrix} \quad (17)$$

Here, $\mathbf{S}(k)$ and $\mathbf{S}(-k)$ represent the ordered set of complex amplitudes $\{S_1(k),...,S_N(k)\}$ and their complex conjugate $\{S_1(-k),...,S_N(-k)\}$. Equation (17) expresses the fact that the transformation of the spin configuration by the operation $(R,\theta|t)$ can compensated by a phase shift $\tau$ such that the spin configuration is kept invariant.

The invariance equation (17) can be used to derive all possible different superspace symmetries resulting from the condensation of all possible types of single-k magnetic orderings described by a single magnetic irrep. For the case of non-magnetic distortions this problem has been systematically analysed [18-21], and the set of all possible (3+1)-dim superspace groups resulting from a single active irrep were calculated and listed in [21]. These superspace groups are obtained as *isotropy subgroups* of the continuous symmetry group associated with the parent structure by adding to the conventional space group operations the continuous set of global phase shifts of the modulation. A complete list of these non-magnetic *isotropy* superspace groups can also be found in the ISOTROPY webpage [28]. We will see in section 3.4 how the possible (3+1)-dim magnetic superspace groups resulting from a magnetic ordering with symmetry properties given by a single irrep can be easily obtained from these lists of non-magnetic superspace groups.

*3.3 Superspace symmetry and irreducible representations*

For a magnetic irrep $mT$, and for an ordered basis of the irrep subspace spanned by the vectors $k$ and $-k$, such that its amplitudes are ordered in the form $\{S_1(\mathbf{k}),...,S_N(\mathbf{k}),S_1(-\mathbf{k}),...,S_N(-\mathbf{k})\}$ (hereafter referred to as a *conjugate ordered* basis), the matrix $mT(R,\theta\,|\,t)$ associated in equation (17) to an operation $(R,\theta\,|\,t)$ belonging to $\Omega_{p,k}$ can be expressed as:

$$\begin{pmatrix} \theta\,\mathbf{D}_T(R)e^{i2\pi k.t} & \mathbf{0} \\ \mathbf{0} & \theta\,\mathbf{D}_T^*(R)\exp^{-i2\pi k.t} \end{pmatrix} \quad (18)$$

where $\mathbf{D}_T(R)$ denotes a $N \times N$ matrix associated with $R$ and $\mathbf{0}$ is the null $N \times N$ matrix. The operation $R$ belongs to the so-called little co-group, a point group formed by all point-group operations present in the elements of the little group $\Omega_{p,k}$. The matrices $\mathbf{D}_T(R)$ form, in general, a *projective* irreducible representation of the little co-group [29], which fully determines both irreps $T$ and $mT$. The $N \times N$ matrices $\theta\,\mathbf{D}_T(R)e^{i2\pi k.t}$ form an irrep of the little group $\Omega_{p,k}$ (*small irrep*), which is sufficient to generate the irrep $mT$ of the extended little group $\Omega_{p,k,-k}$. Except for incommensurate wave vectors at the border of the Brillouin zone in non-symmorphic space groups, the representation $\mathbf{D}_T(R)$ is an ordinary irreducible representation of the little co-group [29]. The magnetic character of the irrep $mT$ is taken into account by the factor $\theta$ multiplying the matrix $D_T(R)$ in (18), so that the matrices of the operations that include time reversal are just the opposite of the corresponding operation without time reversal. The first diagonal matrix block in (18) acts

Magnetic superspace groups and incommensurate magnetic phases on the amplitudes $\{S_i(\boldsymbol{k})\}$, while the second matrix block acts on their complex conjugates $\{S_i(-\boldsymbol{k})\}$. The two blocks are, by definition, related by complex conjugation.

In the case of the operations $(\boldsymbol{R},\theta\,|\,\boldsymbol{t})$ that belong to the coset $g_{-\boldsymbol{k}}\Omega_{p,\boldsymbol{k}}$, and for a conjugate ordered basis, as defined above, the irrep matrices $m\boldsymbol{T}(\boldsymbol{R},\theta\,|\,\boldsymbol{t})$ have the form:

$$\begin{pmatrix} \boldsymbol{0} & \boldsymbol{A} \\ \boldsymbol{A}^{*} & \boldsymbol{0} \end{pmatrix} \qquad (19)$$

with $\boldsymbol{A}$ being a $N \times N$ matrix dependent on the particular operation. It is sufficient to know the this matrix for the chosen coset representative $g_{-\boldsymbol{k}}$ to derive the matrices for the rest of elements of the coset, by multiplying with the matrices of type (18) corresponding to the elements of $\Omega_{p,\boldsymbol{k}}$.

For multidimensional small irreps ($N > 1$), the solution of (17) depends in general on the specific direction taken by the $N$-dimensional vector $\{S_1(\boldsymbol{k}),....,S_N(\boldsymbol{k})\}$. Therefore, several different superspace symmetry groups are in principle possible for the same irrep. Each complex component of the vector $\{S_1(\boldsymbol{k}),....,S_N(\boldsymbol{k})\}$ has its own phase, while there is only a single global shift $\tau$ in (17) to play with. In general, not all operations of the extended little group $\Omega_{p,\boldsymbol{k},-\boldsymbol{k}}$ are maintained in the superspace group and each case has to be considered separately. Therefore, the specification of a given irrep, is clearly insufficient to specify the symmetry of the incommensurate phase and the limitations of a representation analysis without additional symmetry considerations become evident.

On the other hand, for irreps with one-dimensional small irreps ($N = 1$), there is a one-to-one relationship between a given irrep of the paramagnetic space group and a superspace group. For $N=1$, the matrices (18) and (19) are two dimensional and the spin wave amplitudes reduce to two complex conjugated components $\{S(\boldsymbol{k}), S(-\boldsymbol{k})\}$. In this case the values of $D_T(\boldsymbol{R})$ in (18) are either +1 or -1 (for polar axial groups they can also be complex factors, but similar conclusions are obtained, with phase shifts of type 1/3, 1/4 or 1/6, instead of 1/2), and therefore (17) is fulfilled for all operations $(\boldsymbol{R},\theta\,|\,\boldsymbol{t})$ of $\Omega_{p,\boldsymbol{k}}$, with a phase shift $\tau = -\boldsymbol{k}\cdot\boldsymbol{t}$ if $\theta \cdot D_T(\boldsymbol{R}) = +1$ or $\tau = -\boldsymbol{k}\cdot\boldsymbol{t} + \tfrac{1}{2}$ if $\theta \cdot D_T(\boldsymbol{R}) = -1$. Hence, considering that $\tau_0 = \tau + \boldsymbol{k}\cdot\boldsymbol{t}$, all the operations $(\boldsymbol{R},\theta\,|\,\boldsymbol{t})$ of $\Omega_{p,\boldsymbol{k}}$ will become part of the superspace group of the system, either as operations $\{\boldsymbol{R},\theta\,|\,\boldsymbol{t}\,0\}$ or $\{\boldsymbol{R},\theta\,|\,\boldsymbol{t}\,\tfrac{1}{2}\}$. Similarly, the operations that transform $\boldsymbol{k}$ into $-\boldsymbol{k}$ (the coset $g_{-\boldsymbol{k}}\Omega_{p,\boldsymbol{k}}$) will also satisfy equation (17). This can be shown by considering the coset representative $g_{-\boldsymbol{k}} = (\boldsymbol{R}_{-\boldsymbol{k}},+1\,|\,\boldsymbol{t})$. If the small irrep is one dimensional, the form of A in eq. (19) can only be either $+e^{i2\pi \boldsymbol{k}\cdot\boldsymbol{t}}$ or $-e^{i2\pi \boldsymbol{k}\cdot\boldsymbol{t}}$. It is then obvious that equation (17) is satisfied either with

$\tau = -\mathbf{k} \cdot \mathbf{t} + 2\phi$ or $\tau = -\mathbf{k} \cdot \mathbf{t} + 2\phi + \frac{1}{2}$, with $\phi$ being the phase of the complex amplitude $S(\mathbf{k})$. Hence, either $\{R_{-k},+1 | \mathbf{t}\ 2\phi\}$ or $\{R_{-k},+1 | \mathbf{t}\ \frac{1}{2}+2\phi\}$ is a superspace group symmetry operation of the system (the shift along the internal space of the operation depends on the choice of origin along the internal space and can be made zero). The group structure and decomposition (16) then guarantees that all elements of $g_{-k}\Omega_{p,k}$ will be maintained as elements of the superspace group. Summarizing, single-k incommensurate magnetic orderings according to one single irrep with a one-dimensional small irrep, always maintain in the superspace symmetry all operations of the extended little group $\Omega_{p,k,-k}$. A translation (000½) along the internal space is added for operations whose point-group part has character -1 in the small irrep, and no internal translation is added for those with character +1. The internal translations to be added to the operations of the coset $g_{-k}\Omega_{p,k}$ are directly derived considering the internal product of the group, and the fact that no internal translation is necessary for the coset representative $g_{-k}$. This result is very important when considering possible multiferroic properties. It implies that such type of incommensurate magnetic orderings will never break the magnetic point group associated with the extended little group of $\mathbf{k}$, $\Omega_{p,k,-k}$. If the paramagnetic space group contains space inversion, this symmetry operation will necessarily be maintained. More generally, if the paramagnetic phase is non-polar, one can generally say that a magnetic ordering according to an irrep with a 1-dim small representation can never break the symmetry into a polar one, and therefore can never induce ferroelectricity.

*3.4 Time reversal plus phase shift of the modulation as symmetry operation*

Let us consider more closely the consequences of the presence of time reversal as a symmetry operation of the paramagnetic group. As any irrep corresponding to a magnetic order parameter associates the inversion matrix $-\mathbf{1}$ to the time reversal operation $(1'|0\ 0\ 0)$, it is obvious from (17) that the operation $(1'|000,\frac{1}{2})$ will necessarily belong to the superspace group. In fact, this is a general property of any single-k incommensurate magnetic modulation, as it is the consequence of the harmonic character of any primary magnetic arrangement. It is clear that for a harmonic wave, a phase shift of $\pi$ changes the sign of all local magnetic moments. Therefore, the combined action of this phase shift with a time reversal necessarily keeps the system invariant.

This simple general symmetry property has important consequences. It implies that any possible superspace group $\Omega^s$ describing the symmetry of a single $\mathbf{k}$ magnetic incommensurate modulation can be expressed as:

Magnetic superspace groups and incommensurate magnetic phases

$$\Omega^s = G^s + (I'|\, 000, \tfrac{1}{2})G^s \ , \qquad (20)$$

where $G^s$ is a superspace group formed by all the operations $(R, +1 |\, t, \tau)$ that satisfy the invariance equation (17). Therefore, $G^s$ is necessarily one of the superspace groups calculated in [21] and listed in [26], and all possible magnetic superspace groups $\Omega^s$ can be trivially derived from these non-magnetic counterparts through equation 20.

    A second important consequence has already been mentioned in section 2.2. According to Eqs. (5) and (6), the operation $(I'|\, 000, \tfrac{1}{2})$ implies that the spin modulations in single-*k* incommensurate magnetic phases are constrained to odd order Fourier terms, while structural modulations are limited to terms of even order. This means that, if the magnetic modulation becomes anharmonic within the same phase, only *odd* magnetic harmonics are allowed (otherwise the symmetry would be further broken), while the coupling with the lattice can only produce structural modulations with *even* terms, i.e. with $2k$ as primary modulation wave vector. This property is known to happen in many magnetic incommensurate phases (see the example of chromium below), but its origin and validity can only be fully grasped when perceived as a result of a fundamental superspace symmetry operation.

    The symmetry operation $(I'|\, 000, \tfrac{1}{2})$ also implies that the magnetic point group of the phase includes time reversal. Therefore, single-k incommensurate phases cannot be neither ferromagnetic nor ferrotoroidic, i.e. no magnetization or ferrotoroidal moment can appear as an induced secondary weak effect. This result, which can be considered part of the above mentioned restriction of the magnetic configuration to odd-harmonics, illustrates a fundamental advantage of using superspace symmetry concepts, namely the introduction of all the constraints for any degree of freedom of the system, apart from the primary magnetic modulation.

    There has been in the previous literature on magnetic superspace groups [1,5,10] some confusion about the significance of the operation $(I'|\, 000, \tfrac{1}{2})$. In many systems the observed magnetic modulation is often limited to a single harmonic, and its coupling with the lattice is negligible. In such a case, one can reduce the description of the magnetic arrangement to a harmonic magnetic spin wave, which trivially complies with the constraints imposed by the operation $(I'|\, 000, \tfrac{1}{2})$. Therefore, when the model is *a priori* limited to the first harmonic, one can be tempted to consider that the transformation $(I'|\, 000, 0)$ is equivalent to the transformation $(1|\, 000, \tfrac{1}{2})$. Under this viewpoint, the superspace groups $Pn'2_1m'(0\beta0)$ and $Pn2_1m(0\beta0)s0s$ were considered in [5] as equally valid to describe the symmetry of a particular incommensurate magnetic phase with a sinusoidal modulation, because operations such as $(m'_z|\, 000, 0)$ and $(m_z|\, 000, \tfrac{1}{2})$ were considered undistinguishable. However, these two symmetries are not

equivalent when taken as comprehensive symmetry elements of the system, as they imply quite different constraints upon other degrees of freedom. For instance, crystal tensor properties related with magnetism would be quite different. In the first case, with the magnetic point group $m'2m'$, a ferromagnetic component along $y$ would be allowed, while it would be forbidden for the second symmetry (with the point group $m2m$). The correct approach is therefore to consider the two operations as distinct members of the superspace group of the system. The correct superspace group for the system discussed in [5] is therefore $Pn2_1m1'(0\beta0)s0ss$ which, in terms of a coset expansion can be expressed as:

$$Pn2_1m1'(0,\beta,0)s0ss = Pn2_1m(0,\beta,0)s0s + \left(1'|000,\tfrac{1}{2}\right)Pn2_1m(0,\beta,0)s0s, \qquad (21)$$

in agreement with the general expression (20). The magnetic point group of the system is therefore $m2m1'$, i.e. a symmetry that forbids ferromagnetism.

**4. Incommensurate magnetic structures with one irreducible order parameter**

The identification of the magnetic superspace group of a given incommensurate modulation is an efficient and compact way to indicate all the symmetry forced constraints on the degrees of freedom and on the physical properties of the system. As seen above, the possible crystal tensor properties can be immediately derived from the point group symmetry of the superspace group. But, in addition, superspace symmetry also imposes precise restrictions upon the magnetic and structural distortions that are allowed in that phase. This very important advantage of the superspace formalism will be analysed with some detail in this section, with the help of several illustrative examples of magnetic modulation driven by an irreducible order parameter.

It has been argued that the assignment of an irrep to the magnetic distortion is more restrictive or informative than the assumption of a specific magnetic symmetry [30]. This is certainly not true for incommensurate structures if superspace symmetry is used. As it will be shown below, even in the simplest case of a one-dimensional small irrep, the superspace symmetry introduces either stricter or equivalent restrictions, and in the case of multi-dimensional small irreps, the assignment of a superspace group implies the choice of a particular subspace within the space of magnetic basis irrep modes, something that is beyond the method of representation analysis as it is usually applied.

*4.1 The case of one-dimensional small irreps*

4.1.1- The transition sequence in FeVO$_4$

Magnetic superspace groups and incommensurate magnetic phases

Let us consider again the example given in section 2.2, where the paramagnetic phase has the symmetry $P\bar{1}1'$ and the little group $\Omega_{p,k}$ of the propagation vector ($\alpha,\beta,\gamma$) is limited to $P11'$. In this case, only a single one-dimensional magnetic small irrep exists, with character +1 and -1 for the identity and time reversal, respectively. Therefore, according to the general rules previously discussed, the magnetic ordering originated by a single irrep mode necessarily keeps inversion symmetry $\{\bar{1}\,|\,0000\}$ and the time reversal operation $\{1'|\,000\tfrac{1}{2}\}$. This corresponds to the superspace symmetry group $P\bar{1}1'(\alpha\beta\gamma)0s$, which has been described in detail in section 2.2, including the resulting symmetry restrictions on the magnetic and structural modulation. It is illustrative to compare the superspace description for this simple case with that derived from a representation analysis through computer tools such as FullProf (BasiReps) [15], or similar programs [31,32]. In contrast with the superspace symmetry constraints, these codes introduce no conditions on the possible magnetic sinusoidal modulations of atoms at special positions, and allow independent modulations (basis functions) for the two atoms in any pair related by inversion. This is due to the fact that the basis of modes provided by these programs are only symmetry-adapted to the little group of $\mathbf{k}$, $\Omega_{p,k}$ and not to the operations that interchange $\mathbf{k}$ and $-\mathbf{k}$, which in this case are the only ones that restrict the form of an irrep mode. Therefore, if the user does not introduce additional restrictions, the basis functions provided by the usual programs describe an arbitrary spin harmonic modulation, and the inversion symmetry is in general broken. These general unrestricted spin modulations involve at least two irrep modes with the same irrep (there is only one possible irrep!) with some relative phase shift, which breaks the symmetry associated with a single irrep mode.

This simple case is apparently realized in the compound $FeVO_4$, [33]. This material has a paramagnetic phase with space group $P\bar{1}$, and exhibits at low temperatures two incommensurate magnetic phases with a propagation vector along an arbitrary direction. The first phase is non-polar, while the second one exhibits a spontaneous electric polarization. These transitions seem therefore to correspond to the phase sequence :

$$P\bar{1}1' \;\rightarrow\; P\bar{1}1'(\alpha,\beta,\gamma)0s \;\rightarrow\; P11'(\alpha,\beta,\gamma)0s$$

where inversion is lost and ferroelectricity arises only at the second transition, triggered by the condensation of a second magnetic order parameter of the same symmetry. In [33], the intermediate phase was reported as non-centrosymmetric (despite the absence of a spontaneous polarization), but the appropriate phase constraints between the inversion-related Fe atoms to check for the existence of inversion symmetry were not considered [34]. Therefore the most reasonable scenario remains the symmetry sequence depicted above.

### 4.1.2 The incommensurate phase of CaFe$_4$As$_3$

This metallic compound is orthorhombic and has, at room temperature, the symmetry *Pnma* [35], with four independent Fe atoms at Wyckoff positions 4c ( $x$  ¼  $z$ ). At lower temperatures, two magnetic modulated phases have been reported [35]. The first one is stable in the temperature range 90 K <T< 26 K and is incommensurate, with ***k*** = $(0\beta 0)$ (line $\Delta$ of the Brillouin zone) and $0.375 < \beta < 0.39$. The little magnetic co-group of ***k*** is the grey point group *m2m1´*, formed by the symmetry operations {*E*, $m_x$, $2_y$, $m_z$, *1´*, $m_x´$, $2_y´$, $m_z´$}, and the star has two arms (***k*** and *–**k***). The magnetic irreps are classified according to the irreps of the little co-group (see table 2) and there is a one-to one relationship between each irreps and a magnetic space group. These groups, obtained by applying the rules previously discussed, are listed in table 2. It is experimentally observed that the active irrep for the first phase transition of *CaFe$_4$As$_3$* is $m\Delta_1$ [36]. According to table 2, this irrep implies a superspace symmetry *Pnma1´*$(0\beta 0)000s$ for this modulated phase. The symmetry cards for this superspace group are depicted in table 3.

**Table 2.** Irreps of the little co-group *m2m1'* of the $\Delta$-line in the Brillouin zone, which define the four possible magnetic irreps of the magnetic space group *Pnma1´*. In the last two columns the resulting superspace group is indicated by its label and the set of generators. The additional generators $\{1'|000\tfrac{1}{2}\}$ and $\{\bar{1}|0000\}$ are common to the four groups and are not listed. For simplicity, as usual, we use a single label for the irrep of the little co-group and for the corresponding small irrep, full irrep, etc.

| irrep | 1 | $m_x$ | $2_y$ | $m_z$ | 1´ | Superpace group | Generators |
|---|---|---|---|---|---|---|---|
| $m\Delta_1$ | 1 | 1 | 1 | 1 | -1 | *Pnma1'*$(0\beta 0)000s$ | $\{m_x|\tfrac{1}{2}\tfrac{1}{2}\tfrac{1}{2}0\},\{m_z|\tfrac{1}{2}0\tfrac{1}{2}0\}$ |
| $m\Delta_2$ | 1 | -1 | 1 | -1 | -1 | *Pnma1'*$(0\beta 0)s0ss$ | $\{m_x|\tfrac{1}{2}\tfrac{1}{2}\tfrac{1}{2}\tfrac{1}{2}\},\{m_z|\tfrac{1}{2}0\tfrac{1}{2}\tfrac{1}{2}\}$ |
| $m\Delta_3$ | 1 | -1 | -1 | 1 | -1 | *Pnma1'*$(0\beta 0)s00s$ | $\{m_x|\tfrac{1}{2}\tfrac{1}{2}\tfrac{1}{2}\tfrac{1}{2}\},\{m_z|\tfrac{1}{2}0\tfrac{1}{2}0\}$ |
| $m\Delta_4$ | 1 | 1 | -1 | -1 | -1 | *Pnma1'*$(0\beta 0)00ss$ | $\{m_x|\tfrac{1}{2}\tfrac{1}{2}\tfrac{1}{2}0\},\{m_z|\tfrac{1}{2}0\tfrac{1}{2}\tfrac{1}{2}\}$ |

Magnetic superspace groups and incommensurate magnetic phases

**Table 3.** Representative operations of superspace group $Pnma1'(0\beta 0)000s$ described using generalized Seitz type symbols (left column) and symmetry cards as used in the program JANA2006 [8]. The operations with time reversal are obtained by multiplying the first eight operations by $\{1'|000\tfrac{1}{2}\}$, as indicated symbolically in the last row.

| | | | | | |
|---|---|---|---|---|---|
| $\{1\|0000\}$ | $x_1$ | $x_2$ | $x_3$ | $x_4$ | $+m$ |
| $\{2_x\|\tfrac{1}{2}\tfrac{1}{2}\tfrac{1}{2}0\}$ | $x_1+\tfrac{1}{2}$ | $-x_2+\tfrac{1}{2}$ | $-x_3+\tfrac{1}{2}$ | $-x_4$ | $+m$ |
| $\{2_y\|0\tfrac{1}{2}00\}$ | $-x_1$ | $x_2+\tfrac{1}{2}$ | $-x_3$ | $x_4$ | $+m$ |
| $\{2_z\|\tfrac{1}{2}0\tfrac{1}{2}0\}$ | $-x_1+\tfrac{1}{2}$ | $-x_2$ | $x_3+\tfrac{1}{2}$ | $-x_4$ | $+m$ |
| $\{\bar{1}\|0000\}$ | $-x_1$ | $-x_2$ | $-x_3$ | $-x_4$ | $+m$ |
| $\{m_x\|\tfrac{1}{2}\tfrac{1}{2}\tfrac{1}{2}0\}$ | $-x_1+\tfrac{1}{2}$ | $x_2+\tfrac{1}{2}$ | $x_3+\tfrac{1}{2}$ | $x_4$ | $+m$ |
| $\{m_y\|0\tfrac{1}{2}00\}$ | $x_1$ | $-x_2+\tfrac{1}{2}$ | $x_3$ | $-x_4$ | $+m$ |
| $\{m_z\|\tfrac{1}{2}0\tfrac{1}{2}0\}$ | $x_1+\tfrac{1}{2}$ | $x_2$ | $-x_3+\tfrac{1}{2}$ | $x_4$ | $+m$ |
| $\{1'\|000\tfrac{1}{2}\}$ | $x_1$ | $x_2$ | $x_3$ | $x_4+\tfrac{1}{2}$ | $-m$ |
| | ... $\times\{1'\|000\tfrac{1}{2}\}$ | | | | |

The constraints imposed by the symmetry on the magnetic modulation can be derived from equation (5) by taking into account the invariance of the positions of the Fe-atoms under the operation $(m_y | 0\tfrac{1}{2}0)$. These constraints force the magnetic modulation of the Fe atoms to satisfy:

$$M_x(-x_4) = -M_x(x_4),\ M_y(-x_4) = M_y(x_4),\ M_z(-x_4) = -M_z(x_4) \qquad (22)$$

Equation 22 implies that the $x$ and $z$ components of the modulation can have only sine terms in their Fourier series, while only cosine terms are allowed for the $y$ component. According to the experiments, the magnetic modes are aligned along the $y$-axis. Consequently, for a *single* irreducible magnetic spin wave of symmetry $m\Delta_1$, the modulation functions $(M_x(x_4), M_y(x_4), M_z(x_4))$ must have the form $\left(0, M_{y,1c}^i \cos(2\pi x_4), 0\right)$, with $i=1,2,3,4$ labelling the four independent Fe atoms in the reference unit cell. Only four parameters are needed to describe the magnetic structure. Once again, as in the first example, the fundamental symmetry constraint here is not the limitation to cosine functions of the spin modulation (which is due to a convenient choice of the global phase of the magnetic modulation), but the fact that the modulation functions of the four independent Fe atoms must be in phase. This symmetry constraint is counterintuitive as it involves atoms that are symmetry independent in the paramagnetic phase, but it is absolutely necessary in order to restrict the modulation to a mode having the transformation properties of a single irrep. Arbitrary phase shifts between the modulations of the independent Fe atoms imply the superposition of at least four $m\Delta_1$ modes with arbitrary complex amplitudes, and this necessarily breaks the transformation properties that a magnetic configuration driven by a single $m\Delta_1$ mode should have.

As summarized in table 4, the modulation functions of symmetry related atoms can also be determined from (5) for each of the four Wyckoff orbits. The last column in table 4 indicates the restrictions on the spin modulations of any of the four independent Wyckoff orbits of Fe atoms, as obtained in [36] from a conventional representation analysis. The appearance in this mode description of the phase shift of $\frac{\beta}{2}$ for atoms with different positions along the y direction is only a minor nuisance caused by the different parameterization of the modulations (which uses the argument $\boldsymbol{k}\cdot\boldsymbol{l}$ instead of the argument $\boldsymbol{k}\cdot(\boldsymbol{l}+\boldsymbol{r}_\mu)$ adopted in superspace formalism). If this latter is used, this phase shift disappears and, more importantly, the definition of the modulation functions become independent of the choice of the "zero" cell. However, the real important difference between the two descriptions stands in the free relative phases $\Phi_i$ between the modulations of the four independent Fe atoms that are included in this standard representation mode description. This implies the need of seven parameters for describing the structure: four real amplitudes for the four independent Fe sites, plus three phases, since one phase can always be arbitrarily chosen to be zero. In contrast, as shown in table 4, the superspace analysis shows that there are only four free parameters, corresponding to the amplitudes of the four independent modulation functions, since the four modulations of the four Wyckoff orbits are constrained to be in phase. The model refined in [36] does not include this symmetry restriction. This means that the magnetic point group of the reported model is in fact $m2m1'$, i.e. a symmetry polar along *y*, rather than the $mmm1'$ point group symmetry assumed in the article.

**Table 4.** Relation among the modulation functions $M_y(x_4)$ of the magnetic moments along the *y* direction for the atoms of a Wyckoff orbit *4c* within the superspace group $Pnma1'(0\beta0)000s$. In the fourth column, the modulation functions considered in [36] are shown for comparison. $\beta$ is the *y*-component of the incommensurate propagation vector $\boldsymbol{k}=(0\beta0)$ and $\boldsymbol{l}$ stands for a lattice vector of the basic structure labelling a particular unit cell.

| Superspace operation | Position in the basic structure | $M_y(x_4)$ | $M_y(\boldsymbol{k}\cdot\boldsymbol{l})$ |
|---|---|---|---|
| $\{1\|0000\}$ | atom 1: $x$, ¼, z | $M^i_{y,1c}\cos(2\pi x_4)$ | $M_i\cos[2\pi(\boldsymbol{k}\cdot\boldsymbol{l}+\Phi_i)]$ |
| $\{2_y\|0\tfrac{1}{2}00\}$ | atom 2: $-x$, ¾, -z | $M^i_{y,1c}\cos(2\pi x_4)$ | $M_i\cos[2\pi(\boldsymbol{k}\cdot\boldsymbol{l}+\Phi_i+\tfrac{\beta}{2})]$ |
| $\{m_z\|\tfrac{1}{2}0\tfrac{1}{2}0\}$ | atom 3: $x+½$, ¼, -z+½ | $-M^i_{y,1c}\cos(2\pi x_4)$ | $-M_i\cos[2\pi(\boldsymbol{k}\cdot\boldsymbol{l}+\Phi_i)]$ |
| $\{m_x\|\tfrac{1}{2}\tfrac{1}{2}\tfrac{1}{2}0\}$ | atom 4: $-x+½$, ¾, z+½ | $-M^i_{y,1c}\cos(2\pi x_4)$ | $-M_i\cos[2\pi(\boldsymbol{k}\cdot\boldsymbol{l}+\Phi_i+\tfrac{\beta}{2})]$ |

Therefore, inadvertently, the magnetic structural model proposed in [36] for the incommensurate phase of CaFe$_4$As$_3$ is a non-centrosymmetric one. It is interesting to see how large are the deviations of this refined model with respect to the actual symmetry constraints for a *single* $m\Delta_1$ mode structure or, equivalently, for the correct centrosymmetric superspace group symmetry *Pnma*1′(0$\beta$0)000*s*. The reported refined phases (see table 4) are $\Phi_2 = 0.14(3)$, $\Phi_3 = 0.45(3)$, $\Phi_4 = 0.01(4)$, with the choice $\Phi_1 = 0$. Therefore, the deviations from the "symmetric" values 0 or ½ are very small in all cases, close to their standard deviations, except for phase $\Phi_2$.

Again in this example, the differences with the superspace approach originate in the fact that the employed basis functions are not symmetry adapted for the operations interchanging $k$ and $-k$. These symmetry operations are usually disregarded in the representation method applied to incommensurate structures. Atoms belonging to the same Wyckoff orbit in the paraelectric phase, but related by operations that transform $k$ into $-k$ are usually considered to be *splitted* into independent orbits. This assumption is in general not correct and a fully consistent description in terms of irrep basis modes requires to account for relations among these "splitted" atoms that originate in the operations of the coset $g_{-k}\Omega_{p,k}$. In addition, usually the constraints on the basis modes of incommensurate irreps coming from the need to build a single irrep mode are not considered. As we have seen in this example this additional restriction can imply fixed phase relations between the modulations pertaining to atoms that are independent in the paramagnetic phase. The need to extend the usual representation analysis and to consider the symmetry relations associated with a given irrep for operations transforming $k$ into $-k$ has been pointed out and worked out in some recent publications [37-40] by different methods, including a so-called *non-conventional* use of corepresentations [37]. These works were mainly motivated by the need to rationalize the symmetry properties of multiferroic materials, but the extension of the representation method to include these operations is necessary for all incommensurate magnetic structures. In order to do that it is, however, not necessary the use of corepresentations because ordinary irreps define unambiguously the transformation properties of the corresponding magnetic modulation for operations transforming $k$ into $-k$ (even if described with complex amplitudes). Furthermore, as shown in the simple examples above, once the superspace symmetry associated with a given active irrep is identified, this latter is not further required, and the superspace group provides automatically all relevant symmetry constraints, including those coming from the operations transforming $k$ into $-k$, on the magnetic modulation and any other degree of freedom.

4.1.3 Phase II of chromium

Chromium has a bcc structure with a space group $Im\bar{3}m$ in its paramagnetic phase, and exhibits two distinct incommensurate modulated magnetic phases (see [41,42] and references therein). In one of these two phases (hereafter referred to as phase II) the magnetic moments are ordered according to a longitudinal modulation with a propagation vector $(00\gamma)$ (line $\Delta$ or DT in the Brillouin zone), with $\gamma \approx 0.95$. The little group of this vector is $I4mm1'$. The active irrep is $mDT4$ and the corresponding small irrep is one-dimensional (see table 5). This irrep has a star with 6 arms. However, as we are interested in single-k modulations, we will limit our analysis to the sub-space formed by the pair of vectors $\boldsymbol{k}$ and $-\boldsymbol{k}$, and work with the extended little group $\Omega_{p,k,-k}$, which is $I4/mmm1' = I4mm1' + (\bar{1}|000) I4mm1'$ [43].

With the rules discussed in subsection 3.3, the determination of the symmetry of this modulated phase is straightforward. One has just to add a shift $½$ along the internal space to the operations in the little co-group having character -1 for the irrep $mDT4$ (see table 5), and no shift to the coset representative $g_{-k}=(\bar{1}|000)$. This yields the superspace group $I4/mmm1'(00\gamma)00sss$, which has as generators: $\{4_z|0000\}$, $\{m_x|000\tfrac{1}{2}\}$, $\{\bar{1}|0000\}$ and $\{1'|000\tfrac{1}{2}\}$. Notice that the extended little group does not coincide in this case with the full group and, as a result, the superspace group does not contain all the operations present in the paramagnetic phase. This symmetry corrects the superspace group previously assigned in [1] to the phase II of chromium, which, as already mentioned, overlooked the effect of the symmetry operation $\{1'|000\tfrac{1}{2}\}$.

The restrictions on the magnetic and positional structure of the compound that result from the symmetry $I4/mmm1'(00\gamma)00sss$ can be easily derived. In the paramagnetic phase, the single Cr atom per primitive cell is located at the origin and is invariant for all operations of the paramagnetic group. Hence, according to equation (5), the modulation of the corresponding magnetic moment $\boldsymbol{M}(x_4) = (M_x(x_4), M_y(x_4), M_z(x_4))$ must satisfy the relations:

$$(M_x(x_4), M_y(x_4), M_z(x_4)) = (-M_x(x_4), M_y(x_4), M_z(x_4))$$

$$(M_x(x_4 + ½), M_y(x_4 + ½), M_z(x_4 + ½)) = (M_x(x_4), -M_y(x_4), -M_z(x_4)) \quad (23)$$

$$(M_x(-x_4), M_y(-x_4), M_z(-x_4)) = (M_x(x_4), M_y(x_4), M_z(x_4))$$

$$(M_x(x_4 + ½), M_y(x_4 + ½), M_z(x_4 + ½)) = (-M_x(x_4), -M_y(x_4), -M_z(x_4))$$

Magnetic superspace groups and incommensurate magnetic phases

These relations originate in the action of the four generators of the group on the modulation functions. Together, they imply that the *x* and *y* components of the magnetic moments must vanish by symmetry, while the Fourier decomposition of the *z* component must only include cosine odd terms:

$$M_z(x_4) = \sum_{n=odd} M_{zn} \cos(2\pi n x_4) \quad (24)$$

Similar conclusions can be obtained for the possible structural modulations induced through spin-lattice coupling. For instance, a displacement modulation $u(x_4)$ of the atomic positions or a charge ordering modulation $\rho(x_4)$ are subject to equations analogous to (23) but with local transformations complying with those of a polar vector or a scalar field, respectively. This implies that any displacive modulation must correspond to displacements along *z* and can have only even sine Fourier terms, while an induced charge ordering wave can only have cosine even Fourier terms:

$$u_z(x_4) = \sum_{n=even} u_n^z \sin(2\pi n x_4) \quad (25)$$

$$\rho(x_4) = \sum_{n=even} \rho_n \cos(2\pi n x_4) \quad (26)$$

**Table 5.** Irreducible representations of the little co-group $4mm1'$ that define the two irreps of the magnetic space group $Im\bar{3}m1'$ with wave vector $(00\gamma)$, which can be active in chromium. The corresponding irrep matrices for the extended little group $4/mmm1'$ in a conjugate ordered basis are obtained by applying equation (18), and knowing that the matrix **A** in equation (19) associated with the inversion operation $(\bar{1}|000)$ is [1] (1-dim) and [0,1;1,0] (2-dim), for *mDT*4 and *mDT*5, respectively.

| | $E$ | $2_z$ | $4_z$ | $4_z^{-1}$ | $m_x$ | $m_y$ | $m_{xy}$ | $m_{-xy}$ | $1'$ |
|---|---|---|---|---|---|---|---|---|---|
| *mDT*4 | 1 | 1 | 1 | 1 | -1 | -1 | -1 | -1 | -1 |
| *mDT*5 | $\begin{pmatrix}1 & 0\\0 & 1\end{pmatrix}$ | $\begin{pmatrix}-1 & 0\\0 & -1\end{pmatrix}$ | $\begin{pmatrix}-i & 0\\0 & i\end{pmatrix}$ | $\begin{pmatrix}i & 0\\0 & -i\end{pmatrix}$ | $\begin{pmatrix}0 & -1\\-1 & 0\end{pmatrix}$ | $\begin{pmatrix}0 & 1\\1 & 0\end{pmatrix}$ | $\begin{pmatrix}0 & i\\-i & 0\end{pmatrix}$ | $\begin{pmatrix}0 & -i\\i & 0\end{pmatrix}$ | $\begin{pmatrix}-1 & 0\\0 & -1\end{pmatrix}$ |

Hence, as in the conventional representation analysis, the superspace group of phase II of chromium permits only a longitudinal magnetic modulation for this irrep. However through the assignment of the superspace symmetry one obtains the additional information that higher odd-harmonics, with propagation vectors $n\mathbf{k}$ ($n$ odd), are allowed as secondary induced spin waves, as long as they are in phase with the primary longitudinal spin wave. Indeed third order magnetic diffraction satellites have been observed [41,42,44], indicating the existence of a significant third order harmonic of the spin modulation. Similarly, equation (25) imposes that any possible lattice modulation resulting from the spin-lattice coupling must maintain the average position of the Cr atoms and may only develop even order harmonics. These conclusions are also in agreement with the experimental data, which reveal second and fourth order diffraction satellites that have been ascribed to a strain modulation produced by longitudinal atomic displacements [41,42,44]. Given that the global phase of the incommensurate modulation is arbitrary, the restriction of the Fourier series (25) to sine functions, together with (24), express that the relative phase shift between displacive and magnetic modulations, must be $\pm\frac{\pi}{2}$.

The assignment of a superspace symmetry to the phase II of chromium automatically encompasses all the allowed secondary distortions and their constraints. These latter are obtained as symmetry properties, but it is important to realize that they are caused by the restrictions on the possible physical coupling mechanisms that can induce these secondary modulations. For instance, the higher harmonics of the magnetic modulation are the result of coupling terms of type:

$$S^n(\mathbf{k})S(-n\mathbf{k}) + S^n(-\mathbf{k})S(n\mathbf{k}) \qquad (27),$$

which necessarily induces at equilibrium a non-zero amplitude of the $n^{th}$ harmonic, $S(n\mathbf{k})$, proportional to the $n^{th}$ power of the primary order parameter:

$$S(n\mathbf{k}) \propto S^n(\mathbf{k}) \qquad (28)$$

But coupling terms of type (27) are only allowed for n-odd, because they must be invariant for time reversal. Furthermore, equation (28) also implies that the secondary anharmonic modulations must be in phase with the primary harmonic, as in (24). Similarly, the restrictions on the induced structural modulation originated in the spin-lattice coupling can be obtained through the analysis of the symmetry allowed couplings. A modulation of atomic displacements along $z$ with a wave vector $n\mathbf{k}$ ($n$ even) has the form:

$$\mathbf{u}_n(l) = Q(n\mathbf{k})\mathbf{e}_z e^{-i2\pi n\mathbf{k}\cdot l} + Q(-n\mathbf{k})\mathbf{e}_z e^{i2\pi n\mathbf{k}\cdot l} \quad (n \text{ even}), \qquad (29)$$

Magnetic superspace groups and incommensurate magnetic phases with $\mathbf{u}_n(l)$ denoting the displacement of the Cr atom at the $l^{th}$ unit cell, $\mathbf{e}_z$ a normalized displacement vector along $z$ and $Q(-n\mathbf{k}) = Q(n\mathbf{k})^*$. The complex amplitudes $\left(Q(n\mathbf{k}), Q(-n\mathbf{k})\right)$ transform according to irrep DT1 (identity small irrep). Irrep DT1 also describes the transformation properties of $(S^n(\mathbf{k}), S^n(-\mathbf{k}))$ with n-even (although in a different basis). As a consequence, the following coupling term is allowed by symmetry:

$$i\left(S^n(\mathbf{k})\, Q(-n\mathbf{k}) - S^n(-\mathbf{k})\, Q(n\mathbf{k})\right) \qquad (n \text{ even}), \qquad (30)$$

and implies a non-zero equilibrium value of $\left(Q(n\mathbf{k}), Q(-n\mathbf{k})\right)$ in the form:

$$Q(n\mathbf{k}) \propto i\, S^n(\mathbf{k}) \quad (n \text{ even}) \qquad (31)$$

Although (31) is an approximation, the predicted relative phase shift of $\frac{\pi}{2}$ it imposes between the magnetic and the displacive waves is symmetry forced and has a general validity. Notice that the spin-lattice coupling terms of the type of (30) are restricted to n even due to the requirement of time reversal invariance. Also, the absence of transversal displacive modulations in the phase II of chromium can be verified by the impossibility of forming coupling terms similar to (30) involving these displacements and the primary magnetic modulation. Finally, the assignment of a superspace group to phase II of Cr implies establishing symmetry constraints to its crystal tensor properties, either magnetic or non-magnetic. As the point group symmetry is given by the centrosymmetric grey group $4/mmm1'$, ferromagnetism and linear magnetoelasticity are necessarily forbidden.

*4.2 The case of multi-dimensional small irreps*

In the previous section we have seen that in the case of single-k magnetic orderings with a 1-dim small irrep, there is a one-to-one correspondence between each irrep and a superspace group. Thus, the restrictions *on the first harmonic of the magnetic modulation* originated in the superspace symmetry are equivalent to the restrictions imposed by the adapted symmetry mode analysis, if the effect of the symmetry operations that transform $\mathbf{k}$ into $-\mathbf{k}$ were taken into account. For multidimensional small irreps, the two approaches have more fundamental differences, since the one-to-one correspondence between irreps and superspace groups disappears. For multidimensional small irreps ($N > 1$), the solution of (17) depends in general on the specific direction taken by the $N$-dimensional vector $\{S_1(\mathbf{k}), ...., S_N(\mathbf{k})\}$. Therefore, several different superspace symmetry groups are in principle possible for the same irrep.

The different possible superspace groups that can result from a given active irrep with N>1 can be determined by applying (17) without the need to assign any specific microscopic meaning to the components of the order parameter $\mathbf{S}(\mathbf{k})$. Programs like ISODISTORT [27] or JANA2006 [8] do this calculation for any irrep. Once the possible superspace groups for a given active irrep are derived and one of them is assigned to a magnetic phase, the symmetry restrictions on the magnetic modulation and all other degrees of freedom can be directly obtained, as in the previous cases. Let us consider one concrete example.

### 4.2.1 The phase I of Chromium

The phase I of Chromium corresponds to a transversal spin modulation with propagation vector (0 0 γ) that transforms according to the irrep $mDT5$ of $Im\bar{3}m$ (see table 5). This irrep is four dimensional and the order parameter is fully defined by two complex amplitudes $(S_1(\mathbf{k}), S_2(\mathbf{k})) = (S_1 e^{i2\pi\phi_1}, S_2 e^{i2\pi\phi_2})$. The possible superspace groups can be obtained from the analysis of how these amplitudes are transformed under the extended little group $4/mmm1'$ of the vector $\mathbf{k}$ and by applying the invariance equation (17). Let us consider some examples for operations without time reversal since, as seen in section 3, the extension to the operations with time reversal is straightforward. The operation $(2_z | 000)$ transforms $(S_1 e^{i2\pi\phi_1}, S_2 e^{i2\pi\phi_2})$ into $(-S_1 e^{i2\pi\phi_1}, -S_2 e^{i2\pi\phi_2})$. This means that the superspace operation $\{2_z | 000\frac{1}{2}\}$ will always be present for any value of the amplitudes of the order parameter. The operation $(4_z | 000)$ yields $(-iS_1 e^{i2\pi\phi_1}, iS_2 e^{i2\pi\phi_2})$, meaning that the superspace symmetry operation $\{4_z | 000\frac{1}{4}\}$ will be present for configurations of type $(S_1 e^{i2\pi\phi_1}, 0)$ (or similarly, operation $\{4_z | 000\frac{3}{4}\}$ for $(0, S_2 e^{i2\pi\phi_2})$). The inversion $(\bar{1} | 000)$, transforms the order parameter into $(S_2 e^{-i2\pi\phi_2}, S_1 e^{-i2\pi\phi_1})$, hence according to (17) a superspace symmetry operation $\{\bar{1} | 000\ \phi_1 + \phi_2\}$ exists for configurations of the type $(Se^{i2\pi\phi_1}, Se^{i2\pi\phi_2})$. In this way, all special directions in the order parameter space can be explored and their *isotropy* superspace groups derived. Table 6 lists the seven possible magnetic symmetries. The groups $I4221'(00\gamma)q00s$ and $I4221'(00\gamma)\bar{q}00s$ are associated with physically equivalent enantiomorphic spin configurations[5].

---

[5] The two groups are mathematically equivalent by interchanging $\mathbf{k}$ and $-\mathbf{k}$ [25, 28], but we prefer to distinguish the symmetry of the two solutions keeping unchanged the choice of the propagation vector.

**Table 6.** Possible superspace symmetries of an incommensurate magnetic modulation in a $Im\bar{3}m$ structure with propagation vector $\mathbf{k} = (00\gamma)$ and irrep $mDT5$. The restrictions on the form of the order parameter required for each specific symmetry are indicated in the first column. In general, only one direction of the order parameter is shown from the set of equivalent ones, except in the case that the symmetry of different equivalent domains corresponds to enantiomorphic groups. The choice done of the arbitrary global phase of the spin modulation is shown in the 4$^{th}$ column.

| Order parameter | Superspace group | Phase | Generators (besides $\{1'\|000\tfrac{1}{2}\}$) |
|---|---|---|---|
| $(Se^{i2\pi\phi}, 0)$ | $I4221'(00\gamma)q00s$ | $\phi = 0$ | $\{4_z\|000\tfrac{1}{4}\}\ \{2_y\|0000\}$ |
| $(0, Se^{i2\pi\phi})$ | $I4221'(00\gamma)\bar{q}00s$ | $\phi = 0$ | $\{4_z\|000\tfrac{3}{4}\}\ \{2_y\|0000\}$ |
| $(Se^{i2\pi\phi}, Se^{i2\pi\phi})$ | $Immm1'(00\gamma)s00s$ | $\phi = 0$ | $\{2_z\|000\tfrac{1}{2}\}\ \{\bar{1}\|0000\}$ $\{m_x\|000\tfrac{1}{2}\}$ |
| $(Se^{i2\pi\phi}, Se^{i2\pi(\phi+\tfrac{1}{2})})$ | $Fmmm1'(00\gamma)s00s$ | $\phi = -\tfrac{1}{4}$ | $\{2_z\|000\tfrac{1}{2}\}\ \{\bar{1}\|0000\}$ $\{m_{xy}\|000\tfrac{1}{2}\}$ |
| $(Se^{i2\pi\phi_1}, Se^{i2\pi\phi_2})$ | $I112/m1'(00\gamma)00s0s$ | $\phi_1 = -\phi_2$ | $\{2_z\|000\tfrac{1}{2}\}\ \{\bar{1}\|0000\}$ |
| $(S_1 e^{i2\pi\phi}, S_2 e^{i2\pi\phi})$ | $I2221'(00\gamma)00ss$ | $\phi = 0$ | $\{2_z\|000\tfrac{1}{2}\}\ \{2_y\|0000\}$ |
| $(S_1 e^{i2\pi\phi}, S_2 e^{i2\pi(\phi-1/2)})$ | $F2221'(00\gamma)00ss$ | $\phi = \tfrac{1}{8}$ | $\{2_z\|000\tfrac{1}{2}\}\ \{2_{xy}\|0000\}$ |
| $(S_1 e^{i2\pi\phi_1}, S_2 e^{i2\pi\phi_2})$ | $I1121'(00\gamma)00ss$ | ----- | $\{2_z\|000\tfrac{1}{2}\}$ |

As in the case of phase II, the restrictions on the Cr modulations for all the possible alternative symmetries in table 5 can be derived by using the equations discussed in Section 2. These restrictions are summarized in table 7, and figure 2 depicts schematically the form of the magnetic configurations for some of the symmetries. It is illustrative to see the origin of some these restrictions. For instance, the tetragonal superspace groups force the spin wave to adopt an helical configuration. This is due to the fact that operations such as $\{4_z|000\tfrac{1}{4}\}$ force the modulation of an atom at the origin of the basic unit cell to verify the condition:

$$M(x_4 + \tfrac{1}{4}) = 4_z^+ \cdot M(x_4) \tag{32}$$

This condition implies that the *x* and *y* components of the spin modulation must be in right-handed quadrature. Furthermore, equation (32), combined with the relation $M(x_4 + \frac{1}{2}) = -M(x_4)$ forced by the operation $\{1'|000\frac{1}{2}\}$, implies that the *z*-component of the magnetic modulation must be zero. This means that the symmetry only allows transversal modulations. In addition, the operation $\{2_y|0000\}$ requires that

$$M(-x_4) = 2_y \cdot M(x_4). \tag{33}$$

Together with equation (32) this implies that the first harmonic of $M(x_4)$ must be of the form:

$$(M_x^1(x_4), M_y^1(x_4)) = (M_1 \sin(2\pi x_4), -M_1 \cos(2\pi x_4)) \tag{34}$$

with only a free parameter, $M_1$. Similarly, if a third harmonic exists, it must be of the form:

$$(M_x^3(x_4), M_y^3(x_4)) = (M_3 \sin(2\pi x_4), M_3 \cos(2\pi x_4)), \tag{35}$$

with opposite sign correlation of the two components. These relations are then repeated for higher harmonics depending on their parity. The first harmonic is therefore a helical arrangement along *z* axis, with the spins rotating in the xy plane (see figure 2).

There are group-subgroup relations among some of the possible symmetries listed in table 7, implying that some of the constraints are common to some sets of symmetries, while others disappear as the symmetry is lowered. The operation $\{2_z|000\frac{1}{2}\}$ is common to all of the groups and implies that the magnetic modulation function of an atom at the origin must satisfy the condition $M(x_4 + \frac{1}{2}) = 2_z \cdot M(x_4)$. This requirement, together with the condition $M(x_4 + \frac{1}{2}) = -M(x_4)$ imposed by the operation $\{1'|000\frac{1}{2}\}$, restricts the modulations to be transversal even if higher order harmonics are present. The non-centrosymmetric orthorhombic symmetries produce elliptical rotations of the magnetic moments along the propagation direction, with the axes of the elliptical orbit fixed along the *x* and *y* directions (for the $I222 1'(00\gamma)00ss$ symmetry) or the oblique directions (1 1 0) and (-1 1 0) (for the case of $F222 1'(00\gamma)00ss$). It is remarkable that only in the case of a fully arbitrary modulation in the *xy* plane, the *m*DT5 mode produces a polar symmetry.

Magnetic superspace groups and incommensurate magnetic phases

**Table 7.** Symmetry restrictions on the Fourier series describing the modulations of one atom at the origin for the each of the possible magnetic superspace groups listed in table 5. Components not explicitly listed are zero. The cross relations between the amplitudes of sine and cosine terms are indicated symbolically. If the modulations are restricted to sine or cosine terms, a parenthesis with the word is added. If necessary, the restriction in the order-type of the harmonics is also indicated. The general restriction caused by the symmetry operation $\{1'|000\frac{1}{2}\}$ is given in the second row.

| Superspace group | Magnetic $M(x_4)$ $M(x_4+\frac{1}{2}) = -M(x_4)$ odd harmonics | Displacive $u_z(x_4)$ $u(x_4+\frac{1}{2}) = u(x_4)$ even harmonics | Charge/Occupation $\rho(x_4)$ $\rho(x_4+\frac{1}{2}) = \rho(x_4)$ even harmonics |
|---|---|---|---|
| $I4221'(00\gamma)q00s$ | $M_x(\sin/4n+1) = -M_y(\cos/4n+1)$ $M_x(\sin/4n+3) = M_y(\cos/4n+3)$ | $u_z(\sin/4n)$ | $\rho(\sin/4n)$ |
| $I4221'(00\gamma)\bar{q}00s$ | $M_x(\sin/4n+1) = M_y(\cos/4n+1)$ $M_x(\sin/4n+3) = -M_y(\cos/4n+3)$ | $u_z(\sin/4n)$ | $\rho(\sin/4n)$ |
| $Immm1'(00\gamma)s00s$ | $M_x = 0$ $M_y(\cos)$ | $u_z(\sin)$ | $\rho(\cos)$ |
| $Fmmm1'(00\gamma)s00s$ | $M_x(\cos) = M_y(\cos)$ | $u_z(\sin)$ | $\rho(\cos)$ |
| $I112/m1'(00\gamma)s0s$ | $M_x(\cos)$ $M_y(\cos)$ | $u_z(\sin)$ | $\rho(\cos)$ |
| $I2221'(00\gamma)00ss$ | $M_x(\sin)$ $M_y(\cos)$ | $u_z(\sin)$ | $\rho(\cos)$ |
| $F2221'(00\gamma)00ss$ | $M_x(\sin) = -M_y(\sin)$ $M_x(\cos) = M_y(\cos)$ | $u_z(\sin)$ | $\rho(\cos)$ |
| $I1121'(00\gamma)ss$ | $M_x(x_4), M_y(x_4)$ | $u_z(x_4)$ | $\rho$, no condition |

According to table 7, the possible induced displacive structural modulations of an atom at the origin must be longitudinal for all possible symmetries. This is forced by the mutually incompatible constraints imposed by the operations $\{1'|000\frac{1}{2}\}$ and $\{2_z|000\frac{1}{2}\}$ for transversal displacive modulations. In the case of all higher symmetry groups, the additional symmetry operations constrains further the modulation to sine Fourier terms, while in the case of the tetragonal groups, the displacive modulation is restricted to 4n harmonics due to the relation $u(x_4 + \frac{1}{4}) = u(x_4)$ forced by the operation $\{4_z | 000\frac{1}{4}\}$ (or the equivalent relations with translation ¾ ).

Similar to the previous example, the symmetry restrictions on the direction, phase, and possible harmonics of the structural modulations can be traced back to the symmetry constraints on the spin-lattice couplings. If we denote by $Q(2k)$ the complex amplitude of a longitudinal displacive modulation with wave vector $2k$ (see equation (29)), then the lowest order coupling with the order parameter $(S_1(k), S_2(k))$ is given by the symmetry invariant:

$$i\left(S_1(k)S_2(k) Q(-2k) - S_1(-k)S_2(-k) Q(2k)\right) \tag{36}$$

This coupling is similar to that found in phase II for an order parameter of symmetry *m*DT4 (see equation (30)). The difference here is that it is inactive for the special directions of the order parameter corresponding to helical configurations, where either $S_1(k)$ or $S_2(k)$ are zero. According to (36), the amplitude of the induced second harmonic longitudinal modulation is given at first approximation by $Q(2k) \propto i\, S_1(k)S_2(k)$ and therefore this modulation will be zero in an helical phase, in agreement with the conclusion derived directly from the superspace symmetry. This incompatibility of the helical arrangement with a $2k$ induced structural modulation has been occasionally pointed out under particular physical models of the system [45,46]. A comparison of the derivation of this incompatibility in [45] with the one given above is a vivid illustration of the power and simplicity of superspace formalism.

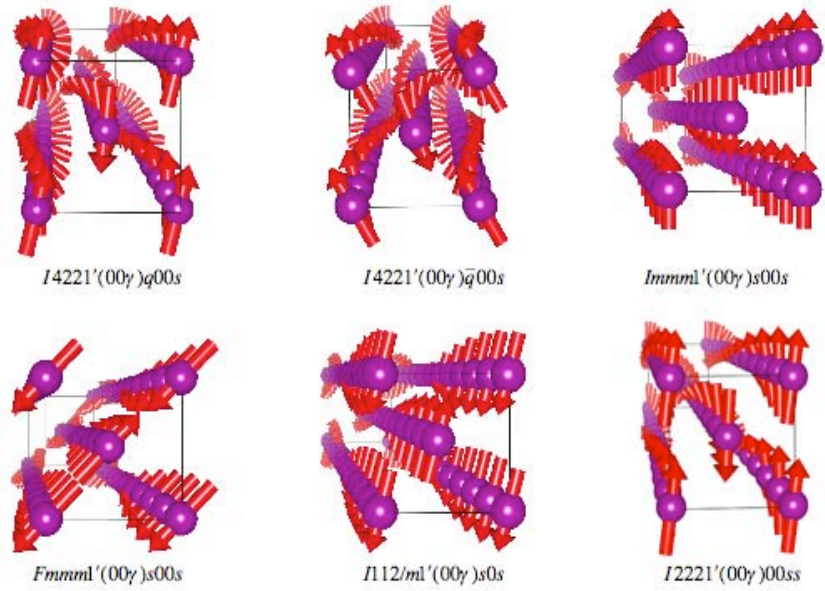

**Figure 2.** Scheme of possible magnetic modes of different superspace symmetry for a bcc structure, with a propagation wave vector (0 0 ~0.96) and irrep $mDT5$. The superspace group correasponding to each case is indicated (see table 6). The figures depict about half wavelength of the incommensurate spin wave. The mode observed in phase I of chromium is the one with superspace group $Immm1'(00\gamma)s00s$.

According to the experimental results, the magnetic moments in phase I of Cr are aligned along either the *x* or *y* direction, with coexistence of both orientations as domains [41]. According to tables 6 and 7, the symmetry of this configuration is given by the orthorhombic group $Immm1'(00\gamma)s00s$. A tetragonal helical arrangement was also proposed in some early works [47], but was later discarded. The experimental distinction between a collinear modulation with equilibrated domain populations and a helical arrangement can be sometimes be difficult, and the possibility of a helical ordering in the phase I of Cr has however persisted in the literature [46, 48]. This contrasts with the symmetry analysis presented above, which shows that a helical arrangement can be directly discarded, since its superspace symmetry is incompatible with the structural modulation with wave vector $2k$ that has been detected in several diffraction studies [41, 42, 44].

4.2.2 Superspace symmetry versus representation analysis for N>1
In the previous example, the assignment of the irrep $mDT5$ to the magnetic ordering only constrains the modulation of the Cr atoms to be a transversal harmonic spin wave of any type. In contrast, each of the possible superspace group symmetries for this irrep restricts the form of the modulation further. Obviously an extended representation method that would specialize the symmetry adapted functions to the special directions in representation space required for each of the superspace groups, would be equivalent to the superspace approach in what concerns the symmetry conditions on the first harmonic modulation. But this complete representation methodology would be unnecessarily complicated, as the most general form of the modulation, including magnetic and structural waves, and any harmonic, can be directly obtained for each special irrep direction from its associated superspace group.

The power of the superspace formalism is that, once a magnetic superspace symmetry is assigned (either derived as a possible one for a certain active irrep, or from inspection of the properties of the experimental data), representation analysis and group theory are no longer needed to describe the structure or its properties. There is no need of building up basis modes, as done in the standard representation method, or to appeal either to the underlying irrep properties of the magnetic ordering. Superspace symmetry operations are defined in an unambiguous form, analogous to space group operations, and the resulting symmetry restrictions on the magnetic modulations and on any other degree of freedom can be directly derived. Then, both the magnetic and the atomic structure can be described (and refined) in a generalized crystallographic manner, considering an asymmetric unit for both the atomic positions and the modulations, with specific constraints on the modulations of the atoms at special positions.

The very particular features of helical structures and other highly regular spin arrangements are usually being introduced by ad-hoc restrictions on the basis irrep modes, when trying to fit their diffraction data [48]. The example above shows that some of the regular features of these arrangements can be assigned to the superspace symmetry of the phase. These features are therefore robust and exact in the sense that their breaking, as it is a symmetry break, requires a thermodynamic phase transition.

## 5. Incommensurate magnetic phases with two active irreducible representations

*5.1 General concepts*

In the previous examples, we have essentially considered possible superspace symmetries of single-k magnetic phases with a unique primary irrep magnetic mode [6]. This implies that the symmetry associated with secondary modes must be, by definition, fully compatible with the symmetry dictated by this primary mode. In the cases discussed above, for example, higher harmonics of the magnetic modulation transforming according to different irreps may occur, but they do not break further the symmetry of the phase, which is solely dictated by the primary mode. However, magnetic phases may also result from the condensation of several primary irrep modes.

---

[6] The term *primary* is used here in the sense that the presence of other (*secondary*) modes within the same phase is explained just as induced or secondary effects.

Magnetic superspace groups and incommensurate magnetic phases

The symmetry of these more general single-k magnetic configurations can be straightforwardly derived by considering the intersection of the superspace groups that would result from each of the primary irrep modes, taken separately. For an experimental example where this type of symmetry analysis has been done, see [12].

Let us then consider a phase that results from the superposition of two irrep primary modes. We will assume that these two modes share a common propagation vector, so that the resulting phase is a single-k magnetic phases describable by a (3+1)-dim superspace group. The superspace groups that may arise from these two modes, taken separately, are not group-subgroup related, and their intersection depends, in general, on the relative phases of the corresponding modulations. As seen in Section 3, in the case of the symmetry operations transforming $k$ into $-k$, the translational part along the coordinate $x_4$ depends on the choice of the origin in the internal space, i.e. it depends on the global phase associated with the modulation. In order to derive the symmetry of the superposition of two active irrep modes, one must then explicitly consider this dependence. When there is an incommensurate modulation with a single irrep, one is always allowed to choose this phase as zero. However, if two primary irrep modulations are superposed, only one of the phases is arbitrary, and the superspace symmetry depends in general on the relative phase shift of the two irrep magnetic modulations.

A shift of the global phase of a modulation by a quantity $\phi$ (in $2\pi$ units) is equivalent to a translation of the origin of the internal coordinate $x_4$ by $-\phi$. Under this origin shift, a symmetry operation $\{R,\theta | t\ \tau_o\}$ becomes $\{R,\theta | t\ \tau_o - R_I \phi + \phi\}$, where $R_I$ is defined in equation (4). This means that the operations that keep $k$ invariant do not change, while those transforming $k$ into $-k$ transform into $\{R,\theta | t\ \tau_o + 2\phi\}$. The intersection of the symmetry groups of the different primary irrep modulations will then depend on their relative phases through their presence in their respective symmetry operations. Let us consider for instance the case of two irrep modes that keep inversion $\{\bar{1} | 0000\}$ in their respective isotropy superspace groups. The translation along the internal space is zero in the two groups, because the global phase of each irrep magnetic mode has been chosen conveniently. However, if the global phases (in $2\pi$ units) of the two modes are $\phi_1$ and $\phi_2$ (with respect to the position of the inversion centre along the internal space), then their inversion symmetry operations are respectively: $\{\bar{1} | 000\ 2\phi_1\}$ and $\{\bar{1} | 000\ 2\phi_2\}$. Hence, a superposition of the two modes will maintain inversion only if $\phi_2 - \phi_1 = n/2$. Similarly, if two irrep modes with a common $k = (0\beta 0)$ are superposed, a first one having a symmetry $\{2_z | 00\frac{1}{2}0\}$ (that is, $\{2_z | 00\frac{1}{2}\ 2\phi_1\}$ for an arbitrary origin in the internal space) and a second one the symmetry $\{2_z | 00\frac{1}{2}\frac{1}{2}\}$ ($\{2_z | 00\frac{1}{2}\ \frac{1}{2} + 2\phi_2\}$, for the same generic origin), then their combined effect will maintain the common two-fold axis only if $\phi_2 - \phi_1 = \frac{1}{4} + \frac{n}{2}$. We have then the necessary ingredients to derive in a straightforward form the possible superspace symmetries produced by the action of two irrep modes with the same propagation vector.

Sometimes ferroelectricity or special magnetoelectric effects originate in complex magnetic orders that involve several primary irreps. We have seen, for instance in section 3, that a single incommensurate irrep magnetic mode with a 1-dim small irrep cannot induce improper ferroelectricity. However, the action of two 2-dim magnetic irrep modes can break the centrosymmetry of a paramagnetic phase and induce a secondary spontaneous polarization, with ferroelectric properties. Therefore the knowledge of the symmetry that results from the presence of several active irreps is especially important for the analysis of possible multiferroic orderings.

*5.2 Multiferroic phases in orthorhombic RMnO3 compounds*

Let us consider the possible irrep magnetic orderings with propagation vector $\boldsymbol{k} = \beta \boldsymbol{b}^*$ in a paramagnetic phase of symmetry $Pbnm1'$ (standard setting $Pnma1'$). This corresponds to the case of the orthorhombic rare-earth manganites of type $RMnO_3$ (R being a rare earth element) [49], which exhibit at low temperatures several modulated magnetic structures with different types of polar behaviour, some of them with two primary irrep modes [50-53].

Table 8 lists the four different possible magnetic irreps of $Pbnm1'$ for a propagation vector $(0\ \beta\ 0)$ and their corresponding superspace groups, according to the general rules explained in section 4. One can then calculate the possible intersections corresponding to the superposition with different relative phase shifts of two primary modes (i.e. configurations of type $m\Delta_i + m\Delta_j$). These possible superspace symmetries are listed in table 9, and can be compared with table 2 in [37], where the non-magnetic point groups of the nuclear structure were listed for the case of two magnetic irreps combined in quadrature. Once taken into account the different settings, the point groups listed there agree with those extracted from table 9. The list in [37] was derived using a so-called "non-conventional application of corepresentation analysis". This reference method indeed considered a non-standard interpretation of the concept of corepresentations. Here, we show that these point groups can be straightforwardly obtained by using *ordinary* irreducible representations of the paramagnetic grey group and their associated superspace symmetries. Moreover, by following the superspace formalism, one obtains not only the point groups to be assigned to the structures, but also the full magnetic symmetry that dictates the restrictions imposed upon any degree of freedom and any tensor property.

The possible ferroic properties of an incommensurate magnetic phase, in particular, are unambiguously defined by the knowledge of its superspace group. From tables 8 and 9, which apply to the $RMnO_3$ compounds, several conclusions can be directly extracted. Firstly, the symmetry operation $\{1'|000\tfrac{1}{2}\}$ is always maintained for phases with two primary irreps.

Magnetic superspace groups and incommensurate magnetic phases

**Table 8.** Irreps of the little co-group *m2m1'* of the $\Delta$-line in the Brillouin zone, which define the four possible magnetic irreps of the magnetic space group *Pbnm1'*. In the last two columns the resulting superspace group is indicated by its label and the set of generators. The generators: $\{1'|000\tfrac{1}{2}\}$ and $\{\bar{1}|0000\}$, common to the four groups, are not listed.

| irrep | 1 | $m_x$ | $2_y$ | $m_z$ | 1' | Superspace group | Generators |
|---|---|---|---|---|---|---|---|
| $m\Delta_1$ | 1 | 1 | 1 | 1 | -1 | $Pbnm1'(0\beta 0)000s$ | $\{m_x|\tfrac{1}{2}0\tfrac{1}{2}0\}, \{m_z|00\tfrac{1}{2}0\}$ |
| $m\Delta_2$ | 1 | -1 | 1 | -1 | -1 | $Pbnm1'(0\beta 0)s0ss$ | $\{m_x|\tfrac{1}{2}0\tfrac{1}{2}\tfrac{1}{2}\}, \{m_z|00\tfrac{1}{2}\tfrac{1}{2}\}$ |
| $m\Delta_3$ | 1 | -1 | -1 | 1 | -1 | $Pbnm1'(0\beta 0)s00s$ | $\{m_x|\tfrac{1}{2}0\tfrac{1}{2}\tfrac{1}{2}\}, \{m_z|00\tfrac{1}{2}0\}$ |
| $m\Delta_4$ | 1 | 1 | -1 | -1 | -1 | $Pbnm1'(0\beta 0)00ss$ | $\{m_x|\tfrac{1}{2}0\tfrac{1}{2}0\}, \{m_z|00\tfrac{1}{2}\tfrac{1}{2}\}$ |

Therefore, ferromagnetism, ferrotoroidicity and linear magnetoelastic or magnetoelectric effects are symmetry forbidden in this type of phases. A second general conclusion is that the superposition of two primary irrep modes that are either in phase or in anti-phase can never induce improper ferroelectricity, because all possible point groups include space inversion. Reversely, space inversion is always broken if the two modes are in quadrature ($\Delta\Phi = \tfrac{1}{4} + \tfrac{n}{2}$), but that does not guarantee the onset of ferroelectricity. As seen in table 9, the combinations in quadrature $m\Delta_1 + m\Delta_2$ and $m\Delta_3 + m\Delta_4$ [7] give rise to the non-polar and non-centrosymmetric point group *222*. For the remaining combinations in quadrature of pairs of modes, the resulting point groups are polar, and therefore an induced ferroelectric polarization is to be expected. The direction of this spontaneous electric polarization depends on the specific pair of irreps. For the combination of distinct irreps, the electric polarization is necessarily oriented along one of the two crystallographic directions perpendicular to the wave vector. This corresponds to the case of the cycloidal spin arrangements observed in the RMnO$_3$ compounds [52-53]. But a polarization parallel to the wave vector is expected for two irrep modes with the same irrep and different global phases. Notice that according to Table 9 only when the two superposed irreps have an arbitrary relative phase shift it is possible an induced polarization along an arbitrary direction in a crystallographic plane. But, even in this case, where the polarization may rotate in the plane as function of temperature or the external magnetic field, a linear magnetoelectric response remains forbidden, due to the presence of the symmetry operation $\{1'|000\tfrac{1}{2}\}$.

---

[7] The symbol $\tau_1 + i\tau_2$ has been sometimes used to indicate the combination in quadrature of two modes with irreps $\tau_1$ and $\tau_2$. This expression can be misleading and is certainly out of the usual notation of group theory.

**Table 9.** Magnetic superspace groups resulting from the superposition of two primary magnetic irreps with a relative phase shift $\Delta\Phi$, for a paramagnetic space group $Pbnm1'$ and a common propagation wave vector $\mathbf{k} = (0\beta0)$ (see table 2 of [37] for comparison).

|  |  | $m\Delta_1$ | $m\Delta_2$ | $m\Delta_3$ | $m\Delta_4$ |
|---|---|---|---|---|---|
| $\Delta\Phi = \frac{1}{4} + \frac{n}{2}$ (Mod.1) | $m\Delta_1$ | $Pb2_1m1'(0\beta0)000s$ | | | |
|  | $m\Delta_2$ | $P2_12_12_11'(0\beta0)000s$ | $Pb2_1m1'(0\beta0)s0ss$ | | |
|  | $m\Delta_3$ | $P2_1nm1'(0\beta0)000s$ | $Pbn2_11'(0\beta0)s00s$ | $Pb2_1m1'(0\beta0)ss0s$ | |
|  | $m\Delta_4$ | $Pbn2_11'(0\beta0)000s$ | $P2_1nm1'(0\beta0)00ss$ | $P2_12_12_11'(0\beta0)0s0s$ | $Pb2_1m1'(0\beta0)0sss$ |
| $\Delta\Phi = \frac{n}{2}$ (Mod.1) | $m\Delta_1$ | $Pbnm1'(0\beta0)000s$ | | | |
|  | $m\Delta_2$ | $P2_1/n1'(0\beta0)00s$ | $Pbnm1'(0\beta0)s0ss$ | | |
|  | $m\Delta_3$ | $P2_1/m1'(0\beta0)00s$ | $P2_1/b1'(0\beta0)0ss$ | $Pbnm1'(0\beta0)s00s$ | |
|  | $m\Delta_4$ | $P2_1/b1'(0\beta0)00s$ | $P2_1/m1'(0\beta0)0ss$ | $P2_1/n1'(0\beta0)s0s$ | $Pbnm1'(0\beta0)00ss$ |
| $\Delta\Phi$ (arbitrary) | $m\Delta_1$ | $Pb2_1m1'(0\beta0)000s$ | | | |
|  | $m\Delta_2$ | $P12_11'(0\beta0)0s$ | $Pb2_1m1'(0\beta0)s0ss$ | | |
|  | $m\Delta_3$ | $P11m1'(0\beta0)0s$ | $Pb111'(0\beta0)ss$ | $Pb2_1m1'(0\beta0)ss0s$ | |
|  | $m\Delta_4$ | $Pb111'(0\beta0)0s$ | $P11m1'(0\beta0)ss$ | $P12_11'(0\beta0)ss$ | $Pb2_1m1'(0\beta0)0sss$ |

It is interesting to consider, within the framework of tables 8 and 9, the properties of the different phases reported for TbMnO$_3$, a most studied representative member of the RMnO$_3$ family. This compound displays a first magnetic phase transition at $T_N \approx 41K$, driven by an active irrep of symmetry $m\Delta_3$. At lower temperatures, $T_C \approx 28K$, a second transition leads to a magnetic phase with a superposition in quadrature $m\Delta_3 + m\Delta_2$ [53]. According to tables 8 and 9, these two consecutive transitions correspond to the symmetry breaking sequence:

$$Pbnm1' \xrightarrow{(T_N)} Pbnm1'(0\beta0)s00s \xrightarrow{(T_C)} Pbn2_11'(0\beta0)s00s \ .$$

The point group of the first magnetic phase is therefore *mmm1'*, and all possible induced structural distortions (restricted to even harmonics) keep space inversion. In the second transition the point group is reduced to *mm21'*, and one should expect an induced secondary polar structural distortion with an electric polarization oriented along *z*.

Magnetic superspace groups and incommensurate magnetic phases

The assigned superspace symmetries not only rationalize the crystal tensor properties of these two phases but, when applied on the possible form of the magnetic modulation, also introduce simple relations between the amplitudes and phases of the spin waves of the symmetry-related magnetic atoms. As some of Tb atoms are only related by operations that exchange ***k*** and –***k***, the symmetry relation between their spin waves is not taken into account by the usual representation analysis. It is remarkable that sometimes these relations have been added, at least partially, with *ad-hoc* arguments. For instance, the amplitudes of the two split Tb orbits were forced to be identical in [53], but their relative phase was refined, when in fact also this phase is symmetry forced.

In the lower temperature magnetic phase of $TbMnO_3$, the magnetic modulation must comply with the superspace group $Pbn2_1'(0\beta0)s00s$; if the magnetic modulation is further restricted to be compatible with A-type local spin arrangements [52], then the reported dominant cycloidal form of the spin modulation [53] is directly obtained from the symmetry conditions of the mentioned superspace group. However, this superspace group also allows the presence of magnetic modulations of type C, F and G. These other types of modulations can introduce in the magnetic modulation complex features beyond the simple cycloidal model and they have indeed been observed, although with weak amplitudes [52, 54].

Under the application of a magnetic field in the yz plane, $TbMnO_3$ undergoes a phase transition in which the polarization rotates from the *z*- to the *x*-axis. According to [55], this transition corresponds to a rotation of the plane of the dominant A-type cycloid. In terms of active irreps, this rotation of 90º of the cycloid plane implies a change of the primary magnetic ordering to a superposition in quadrature of type $m\Delta_3 + m\Delta_1$, which according to table 9, yields the symmetry $P2_1nm1'(0\beta0)000s$, i.e. a phase polar along *x*, with magnetic point group *2mm1'*, explaining the flip of the induced polarization. The above discussion shows that the presence of a spontaneous electric polarization and its orientation, can directly be predicted by symmetry arguments, independently of the microscopic mechanism at work.

## 6. Conclusion

The superspace formalism allows a systematic description and application of the symmetry present in incommensurate magnetic phases. Its relation with the usual representation analysis method has been analysed showing the advantages of a combined use of both approaches. The superspace group defines not only the symmetry restrictions present in the first harmonic of the modulation, corresponding to one or more specific irreps, but it also automatically includes all symmetry restrictions that are present in any other possible induced secondary distortions, such as higher harmonics in the modulated distortion. Magnetic modulated structures are often purely sinusoidal within experimental resolution, and can have a negligible coupling with the lattice, but in the important cases where this coupling is significant (as in multiferroics) and/or the cases where the magnetic modulation becomes anharmonic, the use of the superspace symmetry allows to consider in a systematic way all possible degrees of freedom that, due to the symmetry break, become unclenched.

The magnetic ordering and possible induced structural distortions in an incommensurate magnetic phase are restricted by its superspace symmetry group, and this property is in general more restrictive than the mere description of the magnetic modulation in terms of basis functions for one or several irreps. A consistent comprehensive account of the symmetry properties of single-k magnetic modulations must include its transformation properties for operations changing $k$ into -$k$, and this is done automatically by the superspace symmetry.

We have shown that single-k incommensurate magnetic modulations have the symmetry operation $\{1'|000\frac{1}{2}\}$, combining time reversal and a semi-period phase shift of the modulation. This ubiquitous simple symmetry operation implies important general properties of these systems, as the grey character of their magnetic point groups or the restriction to odd and even harmonics of the magnetic and structural modulations, respectively. To our knowledge, these general symmetry-forced features of single-k magnetic phases, although rather familiar for many experimentalists, seem to have never been formulated in a general context, and their general validity seems to be ignored (see for instance [56]).

An efficient approach to the determination and description of an incommensurate magnetic structure and to the classification of its properties can be achieved by systematically exploring the possible superspace groups associated with one or more irreps, crosschecking successively their adequacy to the experimental data. Recent developments in the programs JANA2006 [8] and ISODISTORT [28] provide tools for the automatic calculation of the possible magnetic superspace symmetries for any paramagnetic space group, any propagation wave vector, and any irrep. This should allow a rapid and systematic exploration in experimental studies of all possible spin configurations, from the highest to the lowest possible symmetries.

The symmetry of *commensurate* magnetic modulations corresponding to the lock-in of the propagation vector into simple rational values (described by conventional Shubnikov space groups) can be directly related with the superspace symmetry of virtual or real neighboring incommensurate phases with irrational propagation vectors. The extreme utility of this close relation between commensurate and incommensurate symmetries is well known in the study of non-magnetic structural modulations. We have not treated here this topic because of lack of space, but some specific examples of its application in magnetic structures can be found in [24]. There it can be seen that, similar to the case of a structural modulation, the magnetic symmetry of a commensurate lock-in magnetic phase depends on the parity of the numerator and denominator of the fraction describing the commensurate wave vector, and well-defined parity rules exist concerning for instance the presence of improper (induced) ferroelectricity. The application of these rules is especially useful to evaluate complex phase diagrams with multiple commensurate and incommensurate phases.

**Acknowledgments**

We thank H. Stokes and B. J. Campbell for valuable discussions and for their readiness to work out and add new features to their software. JMPM gratefully acknowledges helpful comments from L. Chapon, A. Schönleber, S. Krotov and J. Rodriguez-Carvajal. This work has been supported by the Spanish Ministry of Science and Innovation (project MAT2008-05839) and the Basque Government (project IT-282-07). VP thanks Praemium Academiae of the Czech Academy of Sciences for support in developing the program package Jana2006.